\documentclass[twocolumn,showpacs,floatfix]{revtex4}

\usepackage{graphicx}
\usepackage{dcolumn}
\usepackage{epstopdf}
\usepackage{eucal}
\usepackage[dvips]{epsfig}
\usepackage{amssymb}

\usepackage{amsmath,amsthm}

\newcommand{\be}{\begin{eqnarray}}
\newcommand{\ee}{\end{eqnarray}}
\newcommand{\bse}{\begin{subequations}}
\newcommand{\ese}{\end{subequations}}

\newcommand{\bnum}{\begin{enumerate}}
\newcommand{\enum}{\end{enumerate}}

\newcommand{\bit}{\begin{itemize}}
\newcommand{\eit}{\end{itemize}}

\newcommand{\bc}{\begin{cases}}
\newcommand{\ec}{\end{cases}}

\newcommand{\bpm}{\begin{pmatrix}}
\newcommand{\epm}{\end{pmatrix}}

\newcommand{\bvm}{\begin{vmatrix}}
\newcommand{\evm}{\end{vmatrix}}

\newcommand{\bs}{\boldsymbol}

\newcommand{\mcal}{\mathcal}

\newcommand{\mrm}{\mathrm}

\newcommand{\ovl}{\overline}

\newcommand{\ga}{\alpha}
\newcommand{\gb}{\beta}
\newcommand{\gc}{\gamma}
\newcommand{\gd}{\delta}
\newcommand{\eps}{\epsilon}

\newcommand{\gl}{\lambda}
\newcommand{\gk}{\kappa}
\newcommand{\go}{\omega}

\newcommand{\gr}{\rho}

\newcommand{\gs}{\sigma}

\newcommand{\Gd}{\Delta}

\newcommand{\gvf}{\varphi}
\newcommand{\gvr}{\varrho}

\newcommand{\p}{\partial}
\newcommand{\f}{\frac}
\newcommand{\diff}{\mrm{d}}

\newcommand{\lan}{\langle}
\newcommand{\ran}{\rangle}

\newcommand{\kB}{k_B}
\newcommand{\Temp}{\mcal{T}}
\newcommand{\kBT}{k_B\Temp}
\newcommand{\Hyd}{\mcal{H}}

\newcommand{\csp}{\;,\qquad}

\newcommand{\du}{\diff u}

\newcommand{\dX}{\diff X}

\newcommand{\dt}{\diff t}
\newcommand{\ds}{\diff s}

\newcommand{\met}{\mrm{m}}

\renewcommand{\sec}{\mrm{s}}

\newcommand{\kg}{\mrm{kg}}

\newcommand{\Rey}{\mcal{R}}

\begin{document}
\title{CUDA simulations of active dumbbell suspensions}
\author{Victor Putz}
\email{v.putz1@physics.ox.ac.uk}
\author{J\"orn Dunkel}
\email{jorn.dunkel@physics.ox.ac.uk}
\author{Julia M. Yeomans}
\email{j.yeomans1@physics.ox.ac.uk }
\affiliation{Rudolf Peierls Centre for Theoretical Physics, University of Oxford, 1 Keble Road, Oxford OX1 3NP, United Kingdom}
\date{\today}

\begin{abstract}
  We describe and analyze CUDA simulations of hydrodynamic
  interactions in active dumbbell suspensions.  GPU-based parallel
  computing enables us not only to study the time-resolved collective
  dynamics of up to a several hundred active dumbbell swimmers but
  also to test the accuracy of effective time-averaged models.  Our
  numerical results suggest that the stroke-averaged model yields a
  relatively accurate description down to distances of only a few
  times the dumbbell's length.  This is remarkable in view of the fact
  that the stroke-averaged model is based on a far-field
  expansion. Thus, our analysis confirms that stroke-averaged far-field  
  equations of motion may provide a useful starting
  point for the derivation of hydrodynamic field equations.
 \end{abstract}

\maketitle

\section{Introduction}

The derivation of effective hydrodynamic equations from microscopic or
mesoscopic models presents a key problem of non-equilibrium
statistical physics~\cite{2005ToYuRa}. Standard techniques typically involve severe
approximations such as, for example, factorization of correlation
functions, truncation of hierarchies, closure conditions,
etc. Understanding the details of such approximations is crucial for
identifying the range of applicability of the resulting field
equations. If a complex fluid is made up of active
constituents (e.g., bacteria or other micro-organisms) that propel
themselves by quasi-periodic swimming mechanisms~\cite{1977Pu,2009HaMa}, then one usually
faces the additional task of approximating the explicitly
time-dependent microscopic dynamics with a set of coarse-grained, 
time-averaged equations of motion. Aiming at a better quantitative
understanding of this commonly employed approximation, the present
paper provides a detailed comparison between the microscopic dynamics
of actively driven, spring-based dumbbells and those of a
time-averaged analytic model derived from far-field expansion~\cite{2008LaBa,2008AlYe}.
\par
Owing to the fact that hydrodynamic interactions are long-range, simulations of the 
full time-resolved dynamics of $S=N/2$ dumbbells (each consisting of two spheres) 
are numerically expensive, scaling as $N^2$.   However, in the deterministic 
limit case and/or additive noise limit~\cite{1982HaTh}, the dynamics is well suited to parallel 
computations. Very recently, GPU-based codes have been used for various statistical mechanics problems, yielding speed-ups on the order of 20-100 times a CPU-only solution~\cite{2009GeEtAl,2009PrEtAl,2009JaKo,2009ThEtAl}.
Here, we implement a straightforward $N^2$ solution of the hydrodynamic equations of
motion, a communications-intensive task which is difficult to parallelize
in traditional clusters.  For a moderate population size (up to a few thousand particles), 
this method decreases the computation time by a factor of 100 compared with conventional CPU simulations on standard  
consumer hardware.  Hence, we identify GPU computing as a promising approach for future 
 simulations of active particle suspensions (details of the numerical implementation are summarized in Sec.~\ref{s:computational}).
\par
Passive (non-driven) dumbbell models have been widely investigated in
polymer science and related fields over the past decades (see, e.g.,
Refs.~\cite{1969Da,1982TaZa,1989Oet,1993AlFr,2007SeNe,2008BaScZi,2008SuLiGe}).
Very recently, several authors~\cite{2008LaBa,2008AlYe,2009BaMaPNAS}
considered active, internally driven dumbbells as prototype systems
for collective swimming at zero Reynolds number, $\Rey= 0$. Loosely
speaking, one can say that active dumbbell systems constitute a sort
of \lq Ising model\rq~of \emph{collective} swimming, i.e., they
represent strongly simplified models which can be treated by
analytical means, thus providing useful insights.  Active dumbbells
are particularly well-suited to identifying the role of
hydrodynamic long-range interactions in collective micro-swimming.
This is because isolated deterministic dumbbells are prevented from
self-motility by Purcell's scallop theorem~\cite{1977Pu}.  Hence, any
effective motion in deterministic dumbbell systems is caused by
hydrodynamic interactions between different dumbbells.
 \par
 In a recent paper, Alexander and Yeomans~\cite{2008AlYe}
 have derived analytical expressions for the effective far-field 
 interactions of symmetric, active dumbbells in three dimensions (3d).
 Specifically,  they showed that the effective hydrodynamic pair interaction   
 decays with distance~$|D|$ as~$|D|^{-4}$.  Considering  1d motions,  Lauga and Bartolo~\cite{2008LaBa} extended this result to asymmetric
 dumbbells and found that in this case the hydrodynamic interaction decays
  less strongly as $|D|^{-3}$. While these studies have led to novel insights into interplay
between swimmer symmetry and effective long-distance interaction scaling, 
a detailed comparison of microscopic and stroke-averaged models is still lacking. 
The present paper intends to close this gap with respect to symmetric dumbbells.
\par
For this purpose, we shall first introduce a microscopic spring-based
dumbbell model (Sec.~\ref{s:theory}) that can be readily implemented
in GPU-based computer simulations.  In the limit of an infinitely
stiff spring, our model reduces to a shape-driven dumbbell model as
considered in Refs.~\cite{2008LaBa,2008AlYe}. The corresponding 3d
stroke-averaged equations of motion will be discussed in
Sec.~\ref{s:stroke}.  After having confirmed that the
stroke-averaged model reproduces the main features of the microscopic
model simulations in 1d, we perform similar comparative studies for 3d
arrays of symmetric dumbbells. Generally, we find that the
stroke-averaged dynamics yields relatively accurate description of the
microscopic model down to distances of only a few times the dumbbells' 
length. This is remarkable in view of the fact that the stroke-averaged
model is based on a far-field expansion. Thus, at least for the model
considered here, our results suggest that stroke-averaged far-field interaction models 
may indeed provide a useful starting point for the derivation of hydrodynamic field equations.

\section{Microscopic modeling of active dumbbells}
\label{s:theory}

We shall begin by summarizing the ``microscopic" model equations of
the spring-based dumbbells simulated in our computer experiments.  The
corresponding stroke-averaged equations of motion will be discussed
in Sec.~\ref{s:stroke}. To keep the discussion in this part as general
as possible -- and as reference for future work -- we shall formulate
the model for ``Brownian" dumbbells, even though the discussion in the
subsequent sections will be restricted to the deterministic limit.
\par
We consider a system of $S$ identical dumbbells. Each dumbbell
consists of two spheres, of radius $a$.  At very low Reynolds numbers,
inertia is negligible and the state of the system at time $t$ is
completely described by the spheres' position coordinates $\{\bs
X_\ga\}=\{X_{(\ga i)}(t)\}$ with $\ga=1,\ldots, 2S$ labeling the
spheres, and $i=1,2,3$ the space dimension (throughout, we adopt the
Einstein summation convention for repeated lower Latin 
indices).  Neglecting rotations of the spheres, the dynamics is
governed by the Ito-Langevin
equations~\cite{1972MuAg,1976Bi,2009DuLa,2003LiDu,2007SeNe,2009DuZa}
\bse\label{e:langevin}
\be
\dot X_{(\ga i)}(t)
&=&\notag
\sum_\gb \Hyd_{(\ga i)(\gb j)} F_{(\gb j)}
+\\
&&
\sum_\gb (\kB \Temp)^{1/2}C_{(\ga i)(\gb k)}\, \xi_{(\gb k)}(t),
\label{e:langevin-a}
\ee
where $\kB$ denotes the Boltzmann constant and $\Temp$ the temperature
of the surrounding fluid ($\dot
X:=\dX/\dt$). Equation~\eqref{e:langevin-a} corresponds to the
overdamped limit of Stokesian dynamics~\cite{1988BrBo}.  The Gaussian
white noise $\xi_{(\gc k)}(t)$ models stochastic interactions with the
surrounding liquid molecules and is characterized by~\cite{1998GaHaJuMa}
\be
\lan \xi_{(\ga i)}(t) \ran &=&0, \\
\lan \xi_{(\ga i)}(t)  \xi_{(\gb j)}(t') \ran &=&\gd_{\ga\gb}\, \gd_{ij}\,\gd(t-t').
\ee
\ese
The hydrodynamic interaction tensor $\Hyd$ couples the deterministic
force components $F_{(\gb i)}$ that act on the individual
spheres. Generally, the vector $F=\{F_{(\gb i)}\}$ comprises
contributions from internal forces, e.g., those required to bind and oscillate spheres in an active dumbbell, as well as from external force fields (gravity, etc.).
\par
The amplitude of the noise force is determined by the fluctuation
dissipation theorem, which is satisfied if $C$ is constructed from
$\Hyd$ by Cholesky decomposition, i.e.,
\be\label{e:cholesky}
2\Hyd_{(\ga i)(\gb j)}=\sum_\gc  C_{(\ga i)(\gc k)} C_{(\gb j)(\gc k)}.
\ee
In our numerical simulations, $\Hyd$ is given by the
Rotne-Prager-Yamakawa-Mazur
tensor~\cite{1969RoPr,1970Ya,Oseen,HappelBrenner,1982Ma}
\bse\label{e:Mazur}
\be
\label{e:Mazur_diagonal}
\Hyd_{(\ga i)(\ga j)}
&=&
\f{\gd_{ ij}}{\gc_{\ga}}=
\f{\gd_{ ij}}{6\pi \mu a_\ga}
\\
\Hyd_{(\ga i)(\gb j)}
&=&\notag
\f{1}{8\pi\mu\, r_{\ga\gb}} 
\biggl(  \gd_{ij} + 
\f{r_{\ga\gb i} r_{\ga\gb j}}{r_{\ga\gb}^2} 
\biggr)+
\\\label{e:Oseen}
&&
\f{2a^2}{24\pi \mu\; r_{\ga\gb}^3}
\biggl(
\gd_{ij} -
3 \f{r_{\ga\gb i} r_{\ga \gb j}}{r_{\ga\gb}^2} 
\biggr),
\qquad
\label{e:Mazur_offdiagonal}
\ee
\ese
where $r_{\ga\gb i}:=x_{\ga i}- x_{\gb i}$, $\ga\ne\gb$,  and $r_{\ga\gb}:=|\bs x_\ga-\bs x_\gb|$. 
Analytical formulas presented below are based on an Oseen
approximation, which neglects the second line in
Eq.~\eqref{e:Mazur_offdiagonal}.  The diagonal
components~\eqref{e:Mazur_diagonal} describe Stokesian friction in a
fluid of viscosity $\mu$. The off-diagonal
components~\eqref{e:Mazur_offdiagonal} model hydrodynamic interactions
between different spheres.  Note that $\Hyd$ is positive definite for
$r_{\ga\gb}>2a$ and divergence-free, 
$\p_{(\gb j)} \Hyd_{(\ga i)(\gb j)}\equiv 0$ 
with $\p_{(\gb i)}:=\p/\p x_{(\gb i)}$, 
implying that the Cholesky-decomposition~\eqref{e:cholesky} is well-defined.
\par
To completely specify the model, we still need to fix the
intra-dumbbell force. To this end, consider the dumbbell~$\gs$, formed
by spheres $\ga=2\gs-1$ and $\gb=2\gs$, and denote its length by
$d^{\gs}(t):=|\bs X_\gb(t)-\bs X_\ga(t)|$. Neglecting external force
fields from now on, we shall assume that the two spheres are connected
by a harmonic spring of variable length $L^\gs(t)$, i.e., $F_{(\gb i)}=-\p_{(\gb i)}U$ where
\be\label{e:swimmer_potential}
U=\sum_\gs U^\gs,
\qquad
U^\gs(t,d^\gs)
=
\f{k_0}{2}\,[d^\gs- L^\gs(t) ]^2,\; \notag
\\
\ee
with $L^\gs(t)=\ell+\gl \sin(\go t +\gvf^\gs)$ denoting the time-dependent equilibrium
length of the spring, and $\ell$ the mean length such that $\ell > 2a +\gl$. The
dumbbell swimmer is called \emph{passive} if the stroke amplitude
$\gl=0$, and \emph{active} if $|\gl|>0$. As discussed below, the phase
parameter $\gvf^\gs$ is important for the interaction between two or
more dumbbells.
\par
For the overdamped description~\eqref{e:langevin} to remain valid, the
driving must be sufficiently slow.  More precisely, we have to impose
that $T_\gc \ll T_0 \ll  T$, where $ T:=2\pi/\go$ is the driving
period, $T_{0}:=2\pi/ \sqrt{k_0/M}$ the oscillator period for a
sphere of mass $M$, and $T_\gc:=M/\gc$ the
characteristic damping time. This restriction ensures that the
dumbbells behave similar to shape-driven swimmers, i.e., $d^\gs \simeq L^\gs(t)$ 
is a useful approximation in analytical calculations.
\par
With the above assumptions, the $N$-particle PDF $f(t,\{x_{(\ga i)}\})$ of 
the stochastic process $ \{X_{(\ga i)}(t)\}$ from Eq.~\eqref{e:langevin} is 
governed by the Kirkwood-Smoluchowski equation
\be\label{e:kirkwood}
\p_t f
=
\sum_{\ga,\gb}\p_{(\ga i)} \Hyd_{(\ga i)(\gb j)}
\left\{
\left[ \p_{(\gb j)} U\right]f+\kBT \p_{(\gb j)} f 
\right\}.
\ee
For time-independent potentials, the stationary solution of this
equation is given by the Boltzmann
distribution,~$f\propto~e^{-U/(\kBT)}$. However, in the remainder, we
shall focus on the deterministic limit case, formally obtained by
putting $\Temp=0$ in Eqs.~\eqref{e:langevin-a}, which is justified for
sufficiently big spheres.

\section{Stroke-averaged hydrodynamic pair interactions} 
\label{s:stroke}

In this part we will summarize the stroke-averaged equations of motion for  the case of 3d symmetric, deterministic
dumbbell swimmers (a detailed derivation, which differs slightly from that in Ref.~\cite{2008AlYe} but yields equivalent results, is given in the Appendix).  In Sec.~\ref{e:comparison}, the dynamics resulting from these effective equations of
motion for the dumbbell positions and orientations will be compared
with numerical simulations of the microscopic model
equations~\eqref{e:langevin}. From now on all consideration refers to
the deterministic limit case.

\subsection{General stroke-averaging procedure}
\label{s:SA}
We characterize each dumbbell by its direction vector 
\bse
\be
\tilde{\bs N}^\gs(t)=\f{\bs X_{2\gs}-\bs X_{2\gs-1}}{|\bs X_{2\gs}-\bs X_{2\gs-1}|},
\qquad
\gs=1,\ldots, S
\ee 
and its  \emph{geometric} center
\be\label{e:geom_center}
\tilde{\bs R}^\gs(t):=\f{1}{2}\left(\bs X_{2\gs}+\bs X_{2\gs-1}\right).
\ee
\ese
Note that for symmetric dumbbells the geometric center coincides with  the \emph{center of hydrodynamic stress}~\cite{HappelBrenner,2009BaMaPNAS}.
\par
The basic idea of the stroke-averaging procedure~\cite{2008LaBa,2008AlYe,2009DuZa} is to focus on the
dynamics of averaged position and orientation coordinates $\bs R(t)$
and $\bs N^\gs(t)$, which are defined by
\bse
\be
\bs N^\gs(t)
&:=&
\f{1}{T}\int_{t-T/2}^{t+T/2}\du\;
\tilde{\bs N}^\gs(u),
\\
\bs R^\gs(t)
&:=&
\f{1}{T}\int_{t-T/2}^{t+T/2}\du\;
\tilde{\bs R}^\gs(u).
\ee
\ese
Here $T=2\pi/\go$ denotes the period of a swimming stroke. By assuming
that $\tilde{\bs N}^\gs(t)$ and $\tilde{\bs R}^\gs(t)$ are slowly
varying functions of time, one can further approximate
\bse\label{e:stroke_app}
\be
\dot{\bs N}^\gs
\simeq
\dot{\tilde{\bs N}}^\gs
\csp
\dot{\bs R}^\gs
\simeq
\dot{\tilde{\bs R}}^\gs,
\quad
\qquad
\qquad
\\
\f{1}{T}\int_{t-T/2}^{t+T/2}
\!\!\!\!\ds\; f(\tilde{\bs N}^\gs(s),\tilde{\bs R}^\gs(s))
\simeq
 f({\bs N}^\gs(t),{\bs R}^\gs(t))
\ee
\ese
for any sufficiently well-behaved function $f$.

\subsection{Stroke-averaged equations of motion}

Using the approximations~\eqref{e:stroke_app}, one can derive from the microscopic
model equations~\eqref{e:langevin} with $\Temp=0$ the corresponding
deterministic,  stroke-averaged, far-field equations of
motion~\cite{2008LaBa,2008AlYe,2009DuZa}, by making the following
simplifying assumptions:
\begin{itemize}
\item[(i)] The dumbbells are  force-free and torque-free\footnote{If the internal
forces required to contract the dumbbell are central forces, then the
force-constraint implies that the torque-free constraint is
automatically fulfilled.} and approximately shape-driven, i.e.,
~\mbox{$d^\gs:=|\bs X_{2\gs}-\bs X_{2\gs-1}|\simeq\ell+\gl \sin(\go t +\gvf^\gs)$}.
\item[(ii)] The dumbbells are slender, i.e., sphere radius $a$ and
stroke amplitude $\gl$ have about the same size, but are much smaller
than the dumbbell's mean length~$\ell$.
\item[(iii)] The ensemble is dilute, meaning that the distance
$D^{\gs\gr}:=|\bs D^{\gs\gr}|:=|\bs R^\gs-\bs R^\gr|$ between
dumbbells $\gs$ and $\gr$ is much larger than~$\ell$.
\end{itemize}
Adopting (i)--(iii) and restricting to two-body interactions, one
finds the effective equations of motion
\bse\label{e:eom}
\be
\dot{R}_{i}^\gs
&=&
\sum_{\gr\ne\gs} J^{\gs\gr}_i,
\qquad
\label{e:eom_a}
\\
\dot{N}_i^\gs
&=&
- (\gd_{ik}- N^\gs_iN^\gs_k)\;
\sum_{\gr\ne\gs} K^{\gs\gr}_k,
\label{e:eom_b}
\ee
where the stroke-averaged hydrodynamic interaction terms to leading
order in $\gl/\ell$ are given by
\be
\notag
J^{\gs \gr}_i
&=&
a\go\,
\sin(\gvf^\gs-\gvf^\gr)\;\f{9}{64}
 \left(\f{\gl}{\ell}\right)^2
\left(\f{\ell}{|\bs D^{\gs\gr}|}\right)^4\times
\\
& &\notag
\bigl\{
N^\gs_{i}
(2s+4qr-10 sr^2) +
\\
& &\;\;\notag
\hat{D}_{i}^{\gs\gr}(1+ 2 q^2-5 s^2-5r^2\\
& &\;\;\;\;\notag
- 20qsr  + 35 s^2 r^2)
\bigr\},
\label{J}
\\
K^{\gs\gr}_k&=&\notag
\go\sin(\gvf^\gs-\gvf^\gr)\;\f{15}{64}
\left(\f{a}{\ell}\right)
\left(\f{\gl}{\ell}\right)^2
\left(\f{\ell}{|\bs D^{\gs\gr}|}\right)^5\times
\\
& & \notag
\;\hat{D}_{k}^{\gs\gr}
 \bigl( 3s +6rq+ 6sq^2 -7s^3 
\\
& & \;\;\;\notag -  21sr^2-
42qs^2r + 63 s^3r^2
\bigr).\\
\label{K}
\ee
Here, the unit vector $\hat{\bs D}^{\gs\gr}:={\bs
D^{\gs\gr}}/{|\bs D^{\gs\gr}|}$ gives the orientation
of the distance vector 
$\bs D^{\gs\gr}=\bs R^\gs-\bs R^\gr$, 
and $s,r,q$ abbreviate the projections
\be\label{e:projections}
s:=\hat D_{j}^{\gs\gr}N_j^\gs,
\quad
r:=\hat D_{j}^{\gs\gr} N_j^\gr,
\quad
q:=N^\gs_jN^\gr_j.
\quad
\ee
\ese
One readily observes the following prominent features: 
(i) The effective translational interactions scale as~\mbox{$\propto|\bs D^{\gs\gr}|^{-4}$}. 
(ii) The effective rotational interactions scale as~$\propto|\bs D^{\gs\gr}|^{-5}$.
(iii) The stroke-averaged interaction terms $J,K$ vanish if the phases
$\gvf^\gs$ and $\gvf^\gr$ differ by multiples of $\pi$~\cite{2008AlYe}.
This illustrates the importance of  phase (de)tuning in the collective swimming
at zero Reynolds number. 

\section{Microscopic vs. stroke-averaged dynamics}
\label{e:comparison}
We next compare the predictions of the stroke-averaged
equations~\eqref{e:eom} with numerical results obtained from CUDA
simulations of the microscopic spring-based dumbbell model from Sec.~\ref{s:theory}. For this
purpose, we first consider 1d aligned dumbbell pairs similar to those studied by Lauga and
Bartolo~\cite{2008LaBa}.  The rotational interaction of two dumbbells will be analyzed in~Sec.~\ref{s:rotation}. 
Finally, we also study the collective motion of 3d grids of dumbbells (Sec.~\ref{s:arrays}).  
In all cases, the swimmers are assumed to be in an infinite body 
of fluid initially at rest, i.e., no additional boundary conditions (periodic or otherwise) are
imposed.

\subsection{Aligned dumbbell pairs}
\label{s:aligned}
\begin{figure*}[ht!]
\centering
\includegraphics[width=0.3\linewidth]{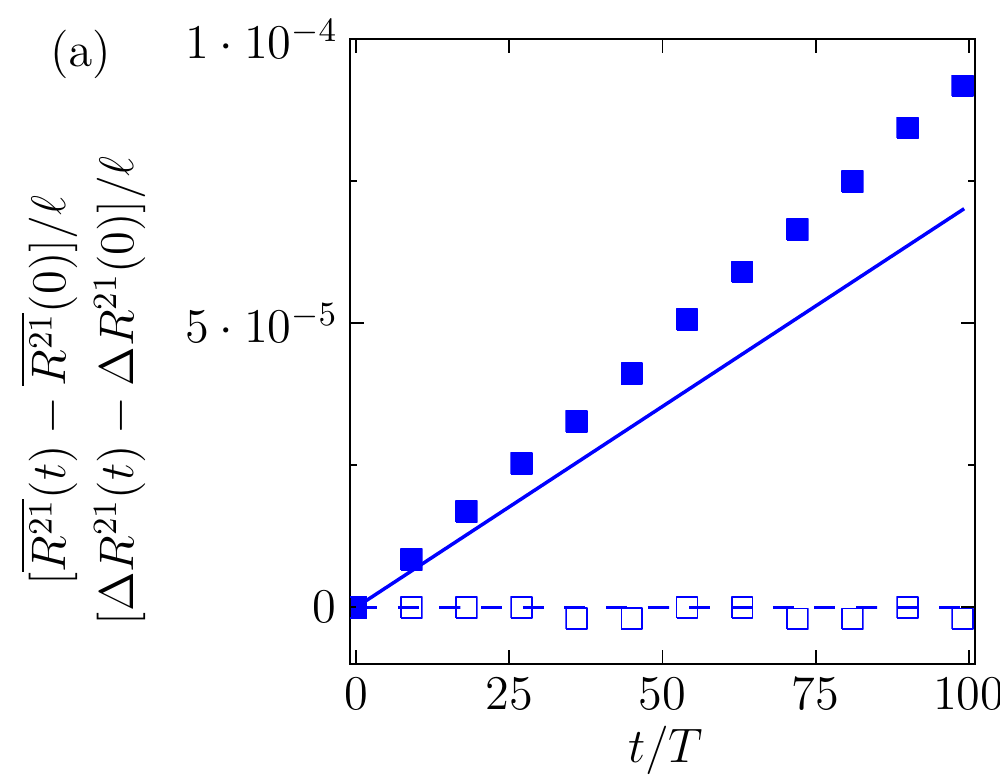}
\includegraphics[width=0.3\linewidth]{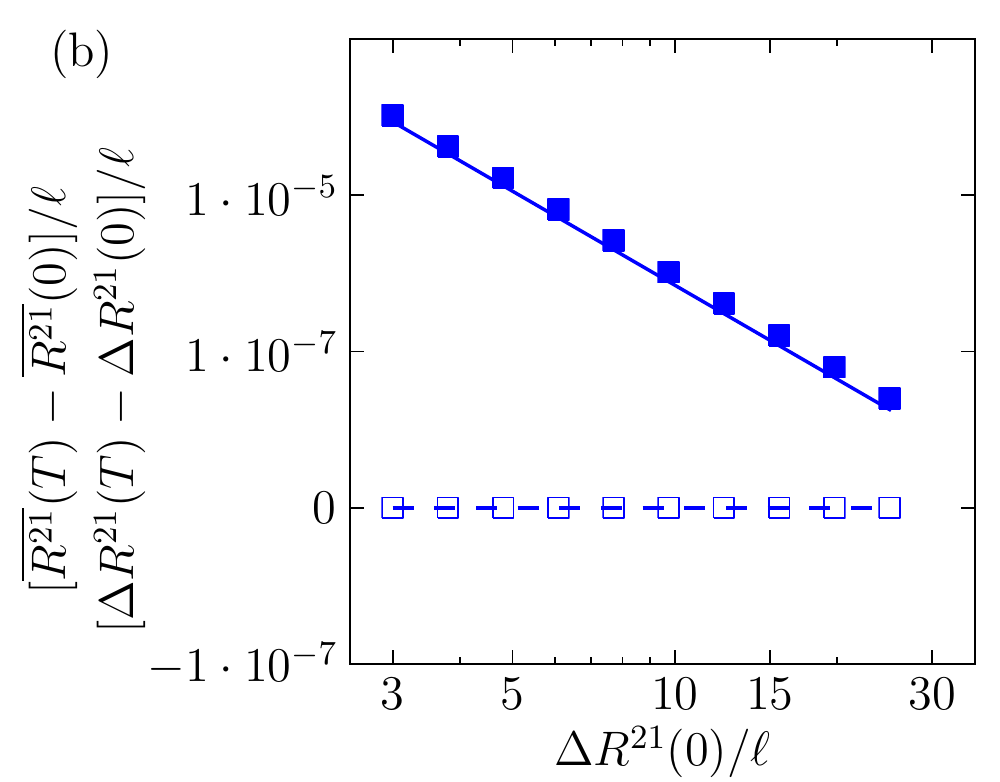}
\includegraphics[width=0.3\linewidth]{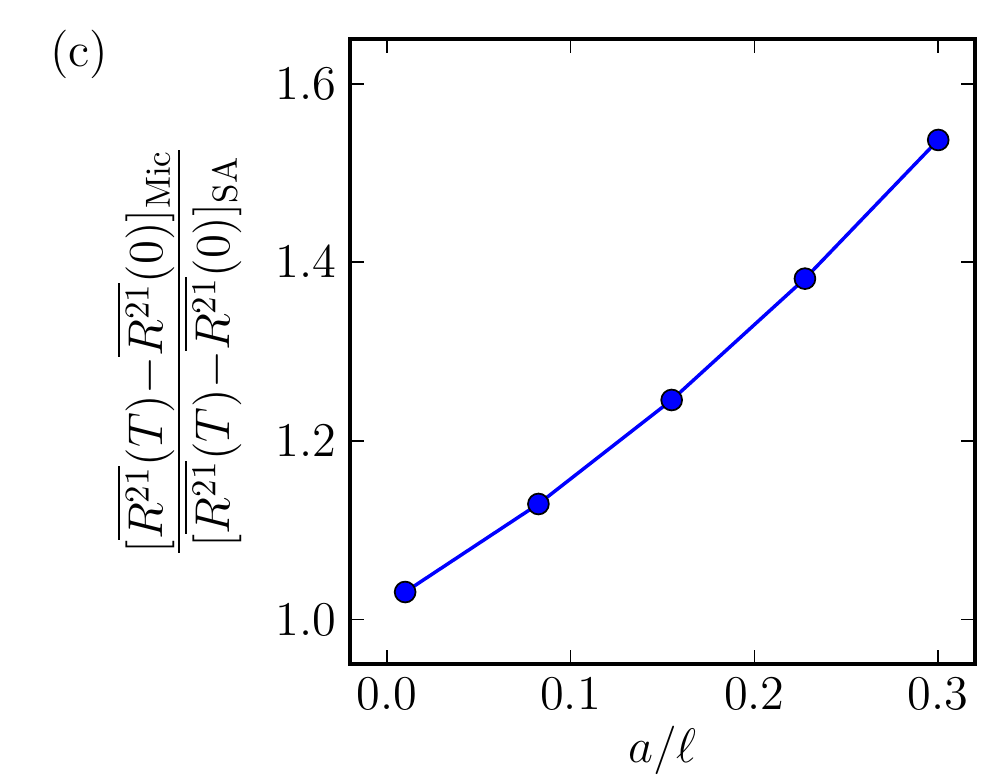}
\caption{
\label{fig01}
Comparison of exact microscopic motion and effective stroke-averaged dynamics  for an aligned dumbbell pair. 
Lines were obtained by numerical integration of the stroke-averaged
equation~\eqref{e:stroke_averaged}, whereas symbols show the simulation results for the
microscopic spring-based dumbbell model described in Sec.~\ref{s:theory}.  Solid lines and filled
symbols depict the mean displacement
\mbox{$\ovl{R^{21} }(t)-\ovl{R^{21} }(0)=\{[R^2 (t)+R^1 (t)]-[R^2 (0)+R^1 (0)]\}/2$}
of the geometric centres. Dashed lines and unfilled symbols indicate
the relative separation \mbox{$\Gd R ^{21}(t)-\Gd R ^{21}(0)=[R^2 (t)-R^1 (t)]-[R^2 (0)-R^1 (0)]$}.
(a)~Symmetric dumbbells do not change their relative separation and move
linearly in time depending on the phase difference $\Gd\gvf=\gvf^2- \gvf^1$.
Simulation parameters are comparable to those of Lauga and
Bartolo~\cite{2008LaBa}: Initial separation $\Gd
R ^{21}(0)=R^2 (0)-R^1 (0)=10\ell$, mean dumbbell
length $\ell=5\mu \met$, driving frequency $\go=500\sec^{-1}$ (time on the x-axis is
given in units of the stroke period $T=2\pi/\go$), stroke amplitude
$\gl=0.1\ell$, $a=0.2\ell$, phase difference $\gvf^2- \gvf^1=\pi/2$.
For the microscopic model: spring constants $k_0=0.001\kg/\sec^2$,
viscosity $\mu=10^{-3}$ kg/(ms), particle mass density
$\gvr=10^{3}$~kg/m$^3$,  simulation time step $\Gd t\approx10^{-4}T$. 
(b)~Distance dependance of the collective motion and separation for
aligned dumbbell pairs during a stroke period~$T$.  Line styles and symbols correspond to the
same configurations and simulation parameters as used in (a).  
Remarkably, the stroke-averaged far-field equation~\eqref{e:stroke_averaged} describes 
the microscopic dumbbell dynamics well down to distances of
a few body lengths; however, the deviations from the time-resolved microscopic dynamics accumulate over
time, as is evident from (a). The difference between the stroke-averaged dynamics (solid lines) 
and the microscopic simulations (symbols) in  (a) and (b) is due to the choice of a relatively large
parameter ratio $a/\ell=0.2$ in these simulations;  the results of both methods agree in the limit $a/\ell\to 0$ 
as illustrated in diagram (c), which shows the ratio of mean swimmer displacements obtained from the  
microscopic  ('Mic') and stroke-averaged ('SA') dynamics at constant $\lambda=0.1\ell$ and various choices of $a/\ell$. }.
\end{figure*}

As long as thermal fluctuations are negligible, aligned dumbbells do
not change their orientation and Eq.~\eqref{e:eom_a} reduces to (see ~\ref{a:SA-1d} for an explicit derivation)
\be
\dot{R}^\gs
=
\f{9}{16} a\go \sum_{\gr \ne \sigma}
\sin(\gvf^\gs-\gvf^\gr)\;
\left(\f{\gl}{\ell}\right)^2
\biggl(\f{\ell}{|D^{\gs\gr}|}\biggr)^4 \hat{D}^{\gs\gr},
\notag\\
\label{e:stroke_averaged}
\ee
where $\dot{R} ^\gs$ denotes the coordinates along the common
axis.  The lines in Figs.~\ref{fig01} (a) and (b) represent the
dynamics of aligned dumbbell pairs ($S=2$) as predicted by
Eq.~\eqref{e:stroke_averaged}. Symbols were obtained from microscopic simulations of 
the corresponding spring-based model described in Sec.~\ref{s:theory}.  Following Lauga and
Bartolo~\cite{2008LaBa}, we quantify collective motion of the dumbbell
pairs in terms of their mean collective displacement (solid
lines/filled symbols in Fig.~\ref{fig01}),
\bse
\be
\ovl{R^{21} }(t)
=
\f{1}{2}[R^2 (t)+R^1 (t)],
\quad
\ee
and their mean relative distance (dashed lines/unfilled symbols in
Fig.~\ref{fig01}),
\be
\Gd R ^{21}(t)
=
R^2 (t)-R^1 (t).
\quad
\ee
\ese
The quantity $\ovl{R^{21} }(t)$ characterizes the net motion of
the dumbbell pair, whereas $\Gd R ^{21}(t)$ indicates whether
the dumbbells the move towards or away from each other. 
\par
As is evident from Fig.~\ref {fig01}~(a), symmetric dumbbells move in the
same direction with identical speeds; the direction of the motion is
determined by the phase difference $\gvf^2- \gvf^1$.  As predicted by
Eq.~\eqref{e:stroke_averaged}, the collective displacement over a
swimming stroke varies as $|D^{\gs\gr}|^{-4}$ with the distance
between the dumbbells, see Figure~\ref{fig01}~(b).  Even though the
stroke-averaged equations~\eqref{e:stroke_averaged} are based on a
far-field expansion, they describe the microscopic dynamics of aligned
dumbbells well down to distances of a few body lengths.
\par
In this context, it is worthwhile to note that the quantitative difference between the 
stroke-averaged dynamics (solid lines) and the microscopic simulations (symbols) in  
Figs.~\ref {fig01}~(a) and (b), is due to the relatively large 
parameter ratio $a/\ell=0.2$ used in these simulations.   As shown explicitly in the Appendix, the stroke averaged equations of motion~\eqref{e:stroke_averaged}  become more accurate in the limit $a/\ell\to 0$. This is confirmed by the numerical results shown in Fig.~\ref{fig01} (c). This diagram depicts the ratio of the average collective  swimming speeds (i.e.,  the collective displacements after a stroke period) obtained by either method for different values of $a/\ell$ at constant stroke-amplitude $\gl$. We readily observe that this ratio approaches unity in the limit $a/\ell\to 0$. 
However, in view of the fact that the collective swimming speed is approximately proportional to the sphere radius $a$, see Eq.~\eqref{e:stroke_averaged}, we opted for a moderate value $a/\ell=0.2$ in all our simulations in order to observe
noticeable swimming effects.

\subsection{Two-dimensional rotation of dumbbell pairs}
\label{s:rotation}
Dumbbells that are arranged in an aligned 1d configuration do not change their orientation. 
This is different for non-aligned configurations in higher dimensions where hydrodynamic pair interactions can induce 
rotations.  To test the accuracy of the stroke-averaged equation~\eqref{e:eom_b}  for the
rotational motions in two dimensions, we conducted a series of
simulations using the following setup:
The first dumbbell (labelled by $\gs$) was placed at the origin oriented
along the $x$-axis, and a second dumbbell ($\gr$) was placed such that the
geometric centres were separated by a distance of $5\ell$.
By varying the starting position of the second dumbbell along a
circle, while keeping the initial projection constant, 
we can compare numeric and analytic results for various 
projections $s(t), r(t), q(t)$, as defined in Eq.~\eqref{e:projections}.
\par
\begin{figure}[t!]
\centering
\includegraphics[width=\linewidth]{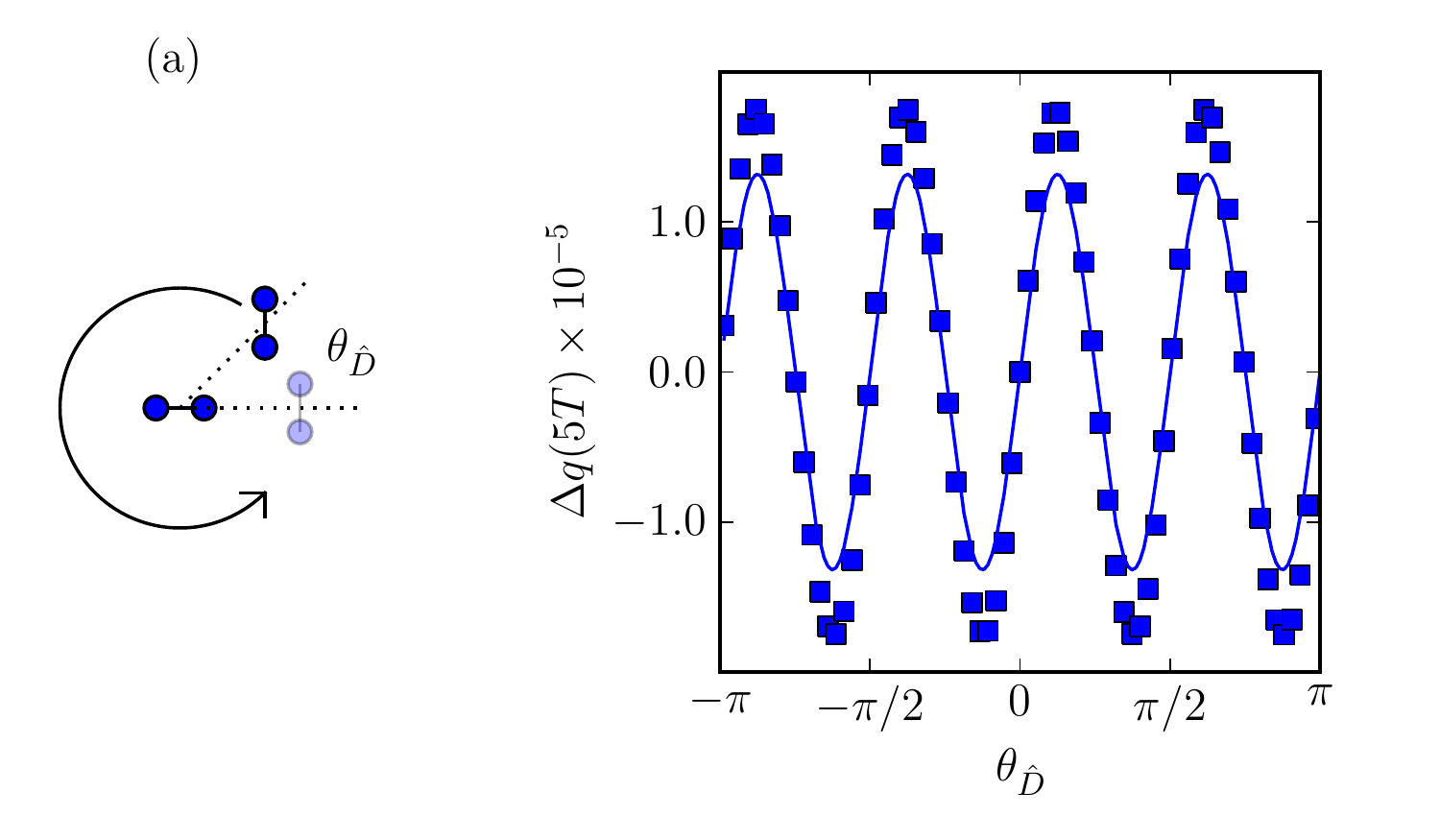}
\includegraphics[width=\linewidth]{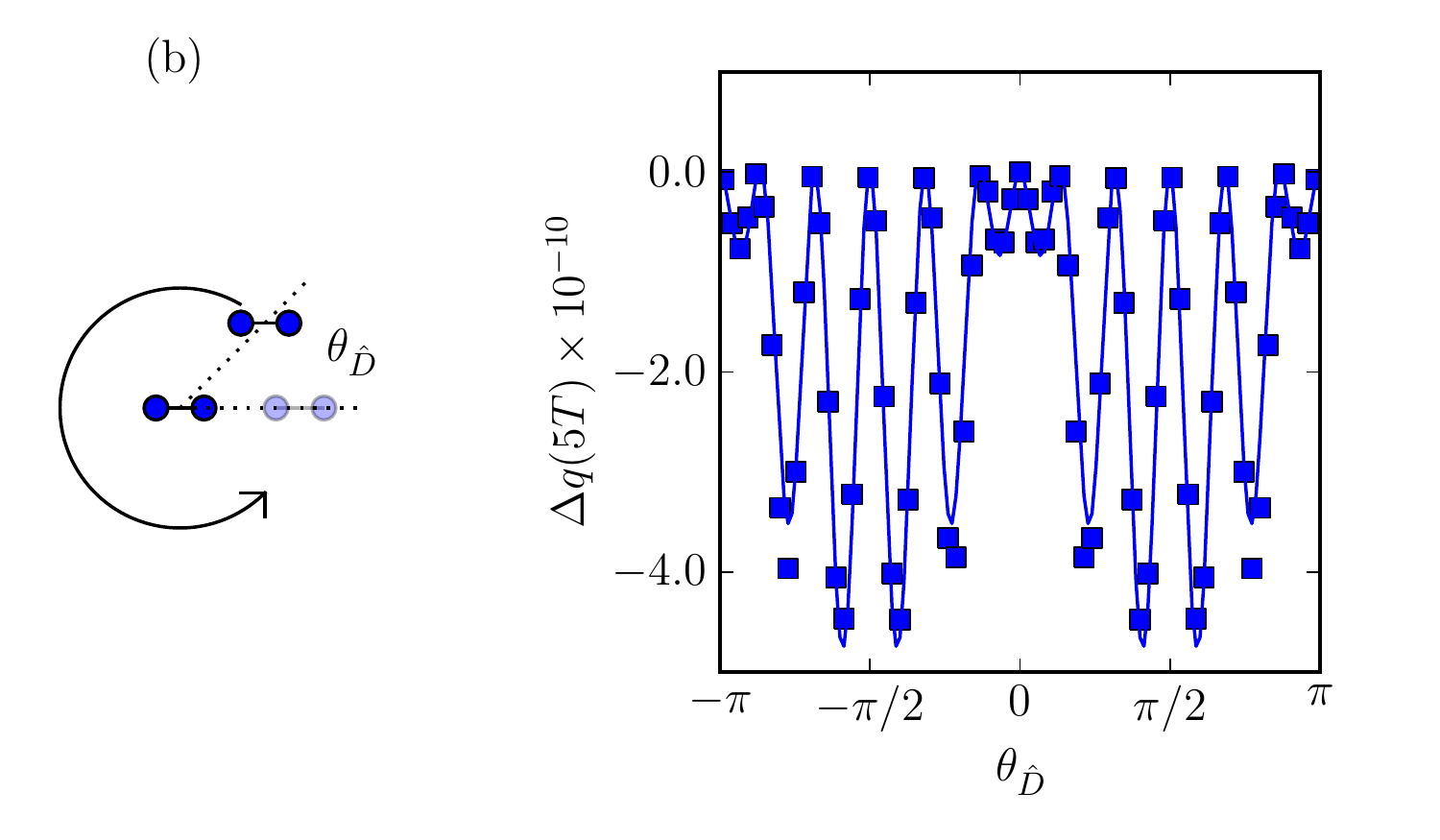}
\caption{
\label{figrotation}
Rotational motion of dumbbell pairs; symbols indicate numeric data
while lines represent analytics.  $\theta_{\hat{D}}$ is the
angle between a line connecting the dumbbell's geometric centres and the 
$x$-axis; $\Delta q(5T):=q(5T)-q(0)$ where $q(t)$ is the projection of the 
swimmer orientations
 $q(t)=N^1_j(t)N^2_j(t)$. Initial configurations:  (a)  $q(0)=0$ and (b) $q(0)= 1$ with an initial radial separation of $5\ell$ . 
 One readily observes the good agreement between the microscopic simulations
and the analytics.
}
\end{figure}

Figure~\ref{figrotation} depicts the change  of the dumbbells' relative orientation
\be
\Delta q(t):=q(t)-q(0)
\csp
q(t):=N^\gs_j(t)N^\gr_j(t)
\ee
after five swimming strokes $t=5T$ for two different initial projections (a) $q(0)=0$ and (b) $q(0)=1$.
As evident from the diagrams, in both cases the stroke averaged description~\eqref{e:eom_b} correctly reproduces the 
rotational dynamics of the microscopic spring-based model.
\par
We may thus briefly summarize: The results in Figs.~\ref{fig01}
and~\ref{figrotation} show that the stroke-averaged
equations~\eqref{e:eom} satisfactorily capture the main features of
effective pair interactions in the spring-based microscopic model at
moderate-to-low densities (large distances). This corroborates that
equations of the type~\eqref{e:eom} can provide a convenient
mesoscopic description which, for example, can be used as a starting
point for derivation of coarse-grained macroscopic field
theories~\cite{2008BaMa}. Conversely, the good agreement between the
averaged dynamics~\eqref{e:eom} and the microscopic model simulation
provides a helpful confirmation that our CUDA algorithm works
correctly even at relatively low densities, when hydrodynamic
interactions effects are relatively weak and algorithms may become
prone to numerical instabilities.
\par
In  the remainder, we shall focus on 3d many-swimmer simulations that fully exploit the virtues
 of the CUDA parallelization scheme.

\subsection{Collective swimming of three-dimensional dumbbell arrays}
\label{s:arrays}

In this section we will compare the predictions of the stroke-averaged far-field
equations~\eqref{e:eom} with simulations of spring-based dumbbells for 3d  dumbbell configurations.
\par
\begin{figure}[t!]
\centering
\includegraphics[width=\linewidth]{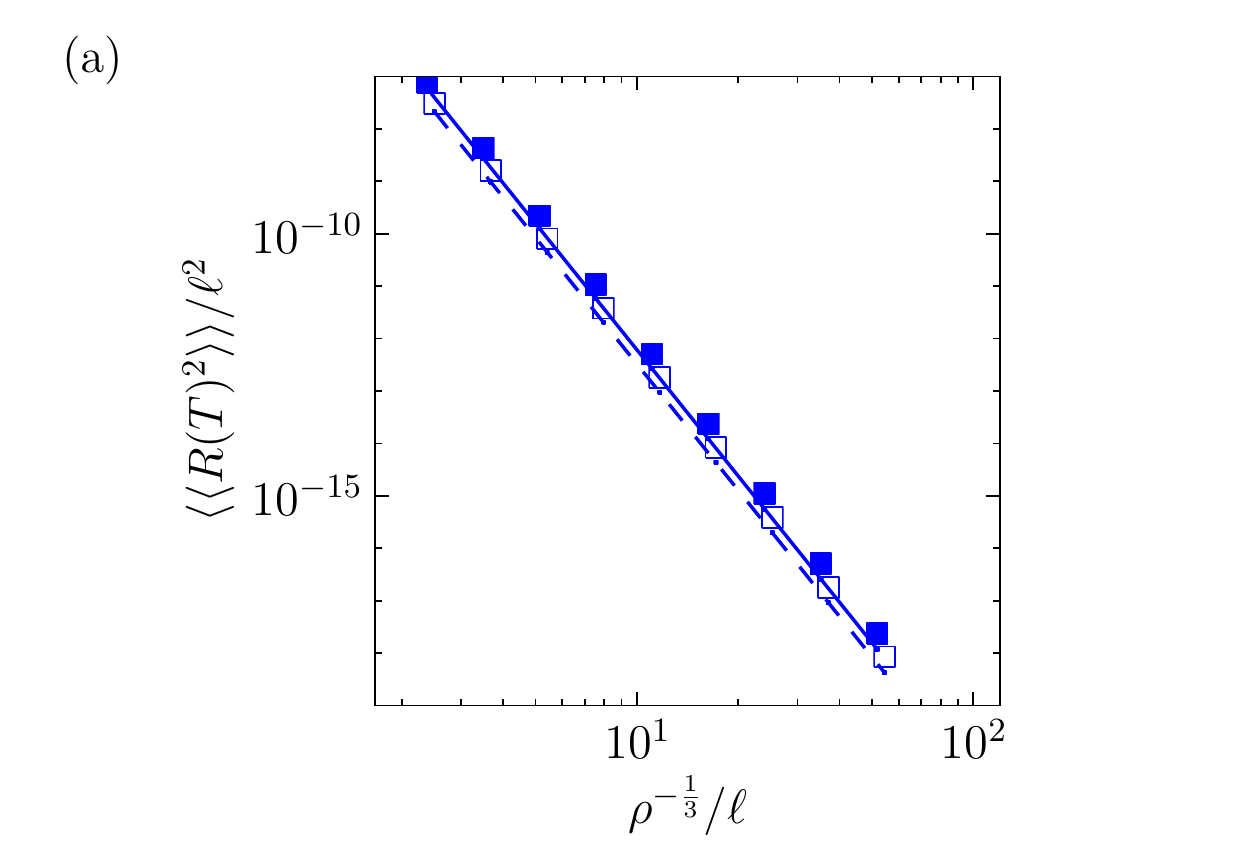}\\
\includegraphics[width=\linewidth]{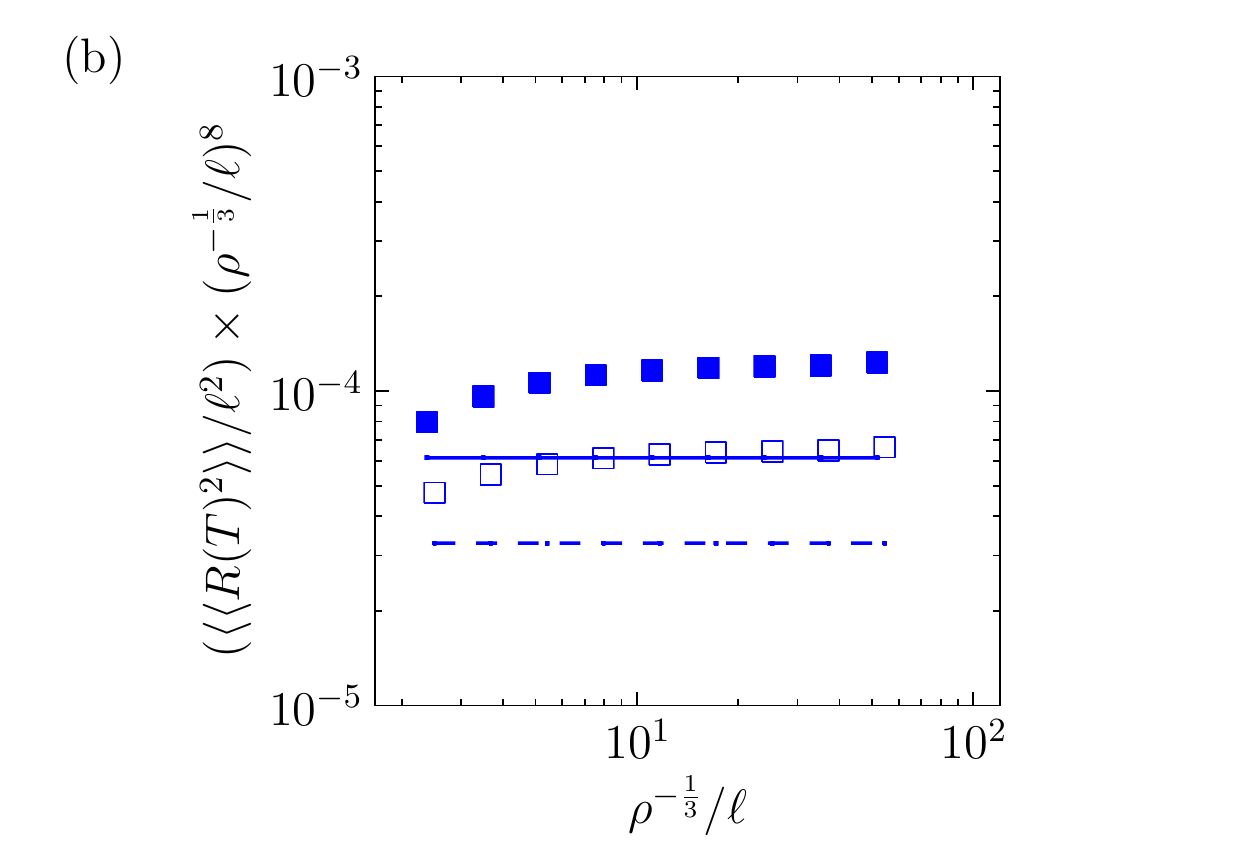}
\caption{
\label{figscaling_with_spacing}
(a) Scaling of the mean square displacement per particle (measured over a
period) with number density for two different phase distributions. The
diagram depicts the simulation results for a cubic array of $(5\times 5\times 5)=125$ 
dumbbells, averaged over $W=100$ different runs, each with
random initial orientations. Symbols refer to the spring-based model and lines
to the stroke-averaged model~\eqref{e:eom}. The collective mean square displacement
is proportional to $(\ell\rho^{1/3})^{2\nu}$ with an exponent
$\nu=-4$.   (b) Mean square displacement rescaled (i.e., multiplied) by $\rho^{-8/3}$.
We observe that for an ``optimized" phase distribution
(filled symbols/solid lines) the effective mean square displacement is
larger than for a uniformly random phase distribution (empty
symbols/dashed lines). On this scale, statistical error bars (not shown) are smaller 
than the size  of the symbols. The shift between lines and symbols,  caused by the relatively large
parameter ratio $a/\ell=0.2$ used in these simulations, is 
consistent with the value expected from Fig.~\ref{fig01}~(c).
 }
\end{figure}
\begin{figure}[t!]
\centering
\includegraphics[width=\linewidth]{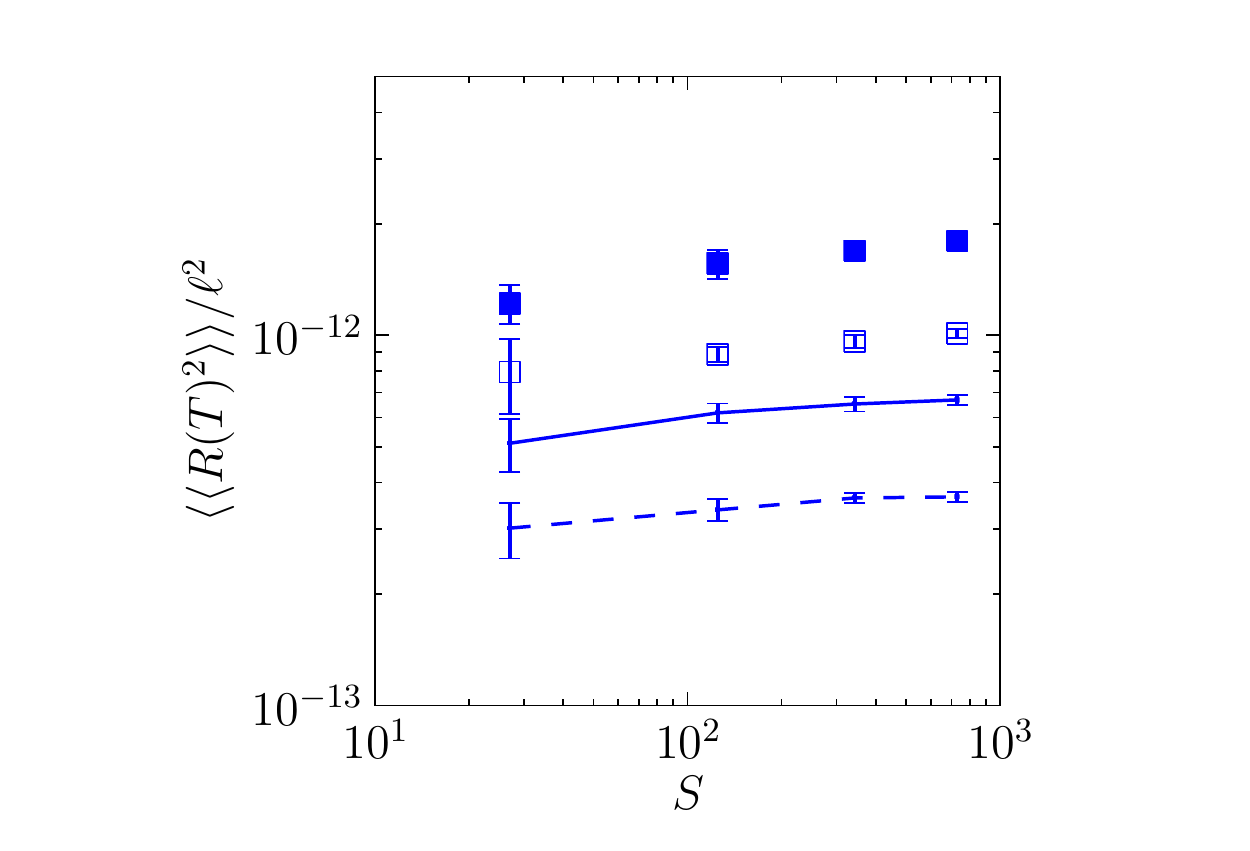}
\caption{
\label{figscaling_with_number}
Scaling of the mean square displacement per particle over a period
with dumbbell number $S$ at constant density. The diagram depicts the
simulation results for collections of dumbbells arranged on a cubic
$(x\times x\times x)$-lattice with $x=3,5,7,9$ and spacing $10\ell$, averaged over $W=10$ different random initial 
orientations. Symbols refer to the spring-based model and lines to the far-field stroke-averaged model~\eqref{e:eom}, beginning from the same initial
conditions. Increasing the number of swimmers while keeping the number density
constant produces only minimal gains in translational speed. 
The collective mean square displacement is smallest for $(3\times 3\times 3)=27$ swimmers, which is  
due to the relatively large number of swimmers with an incomplete set of  ``nearest neighbors'' in this case. 
We also observe that for an ``optimized" phase distribution (filled
symbols/solid lines) the effective mean square displacement is larger
than for a uniformly random phase distribution (empty symbols/dashed
lines). Again, the shift between lines and symbols,  caused by the relatively large
parameter ratio $a/\ell=0.2$ used in these simulations, is 
consistent with the value expected from Fig.~\ref{fig01}~(c).
}
\end{figure}

 In our simulation the dumbbells' geometric centers $\bs
R^\gs(0)$ are initially placed on a cubic $(x\times x\times x)$-lattice 
with equidistant spacing $\gr^{-1/3}$, where $\gr$ is the
number density of the configuration.  Initial orientations $\bs
N^\gs(0)$ are sampled uniformly from the unit sphere.  For the lattice
size we consider values $x=3,5,7,9$ corresponding to a total dumbbell
number $S=3^3, 5^3,7^3, 9^3$, respectively. To characterize the
collective motion, we measure in our simulations the mean square
displacement per particle averaged over different initial conditions,
i.e.,
\be
\lan\lan R(t)^2 \ran\ran:=
\f{1}{W}\sum^W_{w=1}\f{1}{S}\sum_{\gs=1}^S
\left[R^\gs(t;w)-R^\gs(0;w)\right]^2 \notag
\\
\ee
where the variable $w=1,\ldots,W$ labels different initial conditions. 
We distinguish two classes of initial conditions:
\begin{itemize}
\item[(i)] An ``optimized" phase distribution: Phases were set such that each
dumbbell had a phase of $0$ or $\pi/2$ with all nearest neighbors
having the alternate phase in the manner of a 3d ``checkerboard''.
The corresponding results for the microscopic simulation and the
stroke-averaged model are indicated by filled symbols and solid lines
in Figs.~\ref{figscaling_with_spacing} and
\ref{figscaling_with_number}, respectively.
\item[(ii)]
 A randomized phase distribution: Phases were set to random values
evenly distributed on the interval $[0, 2\pi)$, with a different
distribution for each run.  The corresponding results are indicated by
unfilled symbols and dashed lines in
Figs.~\ref{figscaling_with_spacing} and \ref{figscaling_with_number},
respectively.
\end{itemize} 
 
\par 
Fig.~\ref{figscaling_with_spacing} illustrates how the mean square
displacement over a period, $ \lan\lan R(T)^2 \ran\ran$,
varies with density $\gr$ -- or equivalently with grid spacing -- for
an array of $(5\times 5\times 5)=125$ dumbbells.  As evident from the
diagram, the prediction from the stroke-averaged model~\eqref{e:eom}
is in good agreement with the scaling behavior measured for the
microscopic model.  Furthermore, by comparing filled with unfilled
symbols and solid with dashed lines, we note that the collective
displacement $ \lan\lan R(T)^2 \ran\ran$ is generally smaller
for the randomized phase distribution, corroborating the fact that
optimizing the phase distributions can considerably enhance the
effectiveness of collective motions \cite{2008YaElGo}.

\par
Figure~\ref{figscaling_with_number} shows how the quantity $ \lan\lan R(T)^2 \ran\ran$ 
scales with the total number $S$ of the
dumbbells at fixed density~$\gr$.  After a slight initial jump from
the $(3\times 3\times 3)$ case, adding more swimmers at fixed density
$\gr$ produces only a minimal increase in displacement, and the effect
appears for both optimized or randomized phase distributions.
Again, collective displacement is generally smaller for randomized
phase distributions than for optimal phase distributions.

\section{Computational aspects}
\label{s:computational}

The numerical results were obtained from parallelized
simulations run on graphics processing units (GPUs) using Nvidia's
Compute Unified Device Architecture (CUDA).  Compared to 
conventional CPU programs, GPU algorithms may yield significant speed
ups (up to factors of a few hundreds) whenever a problem can be
naturally parallelized~\cite{2009GeEtAl,2009PrEtAl,2009JaKo,2009ThEtAl}, on relatively
low-cost consumer-grade hardware.  In cases where the problem is small
enough to fit on a single device, the resulting software is simpler,
easier to test, less costly to implement, and much faster than
traditional cluster-based methods.  This is the case for deterministic
many-swimmer simulations, for stochastic single-swimmer simulations,
and also for stochastic many-swimmer simulations with purely additive
noise, corresponding to a constant matrix $C$ in
Eq.~\eqref{e:langevin}.
\par
Most $\mathcal{O}(N^2)$ problems such as $N$-body simulations with pair interactions involve
enough data communication that they are difficult to distribute
efficiently, or are costly enough that they must be recast in more
numerically tractable forms such as Ewald 
summation~\cite{saintillan_smooth_2005,brady_stokesian_1988,sierou_accelerated_2001}.  For cases of a
few hundred swimmers, CUDA-based implementations of straightforward
$\mathcal{O}(N^2)$ problems present an excellent method for numerical simulation.
We use a simple direct computation of sphere-sphere interactions via the 
Rotne-Prager-Yamakawa-Mazur tensor, disregarding
lubrication forces and close-range interactions due to the dilute nature
of the suspensions and slender structure of the dumbbells.
\par
We tested our GPU code on an AMD Phenom X4 940 system running Fedora Linux,
using a consumer-level Nvidia GTX 295 GPU and a more research-oriented
Tesla C1060 GPU; other tests took place on an Intel i7 860 running Gentoo
Linux and a consumer-level GTX 276 GPU.  All hardware was capable of double-precision
calculation and used version 2.3 of the CUDA toolkit.
\par 
Initial testing of a similar but simpler problem (colloids moving under
constant applied force, using Oseen interactions and single precision) showed
very large benchmarked speed-ups compared with a C-based CPU simulation. For example, with $\sim2000$ 
particles, we measured up to a $\sim450\times$ speed-up
when calculating the full hydrodynamic interaction tensor and $\sim800\times$ speed-up
using an un-optimized version of the elegant tiled method
described in~\cite{prins_nbody_2007}.
We did not benchmark the current simulation, but estimate the speed-up, while still being significant,
to be considerably less due to the complexity of the multi-swimmer problem, additional data transfers from
the GPU, and the use of double precision.
\par
Despite the speed advantage of the tiled method, we decided to compute 
the full hydrodynamic interaction tensor in our simulations, primarily to maintain congruence with existing
C code in a battery of automated unit tests and to keep the code as simple as 
possible.  Future implementations will likely reintroduce the tiled calculation to  further improve computational efficiency. Generally, it is encouraging that even a relatively straightforward CUDA simulation of the multi-swimmer problem exhibits compelling speed advantages over a CPU-based solution.
\par
The use of double precision is unfortunate in that single precision calculations on
CUDA processors show significant performance increases due to better hardware
support and memory performance.  However, in the case of collective dumbbell motion, 
the distance moved in each time-step is very
small compared to the length of the dumbbells or their position, which caused 
initial calculations using single precision to fail, as the position incremental during a single timestep 
fell below the threshold of machine precision.  To allow for standardized 
testing, we elected to use double precision
and to accept decreased performance rather than implementing a better accumulation 
algorithm (such as Kahan summation) based on single precision.  For reasons of accuracy, 
we also chose not
to enable Nvidia's fast-math optimizations. The latter can significantly accelerate the computation of certain numeric functions (particularly trigonometric functions)  but this gain comes at the cost of some precision.  However,  this might represent another opportunity for performance optimization in the future.
\par
Another important issue is the choice of the integrator due to
accumulation of errors and the stiffness of the problem.  
A variety of methods were tested, including Euler,
Adams-Bashforth-Moulton, and Runge-Kutta integrators.  The approach 
eventually used was a one-step Heun predictor-corrector method, which
produced excellent results and can easily incorporate additive noise
for stochastic simulations.  The time step for the simulations was chosen
based on the smallest dynamical time scale in the problem (given by the spring frequency $T_0=2\pi/\sqrt{k_0/M}$, see discussion in Sec.~\ref{s:theory})  and then
manually reduced until numerical errors were acceptable by ensuring that single dumbbells
did not translate and numeric fluctuations were
orders of magnitude below the expected motion caused due to hydrodynamic interactions.
\par
The use of a spring-based model created an additional
complication: After prescribing the initial position, orientation, and phase of the dumbbell
we initially placed the spheres centred at the potential minima. However, numerical integration and finite potential
strengths caused the sphere positions to lag very slightly behind the
potentials once they began moving periodically.  Since the hydrodynamically induced
dumbbell motion is of a very small scale compared to the dumbbell
size, this initial settling caused a large anomalous motion during the
first period of simulation.  To rectify this, it was necessary to
discard the first period and begin measurements after the lag was established and dumbbell translation was
approximately linear.  This did cause miniscule deviations of the dumbbells' 
mean length $\ell$, amplitude $\lambda$, and phase  $\gvf$ from the values specified
by the initial conditions, but tuning the potential spring constant to be
sufficiently stiff reduced these deviations to acceptable values of a few percent.
\par
While the results shown here are purely deterministic, incorporating
noise is relatively straightforward, as the system hydrodynamic tensor
$\Hyd$ may be numerically decomposed via Cholesky decomposition~\cite{2006Jung,2008VoDe}.
However, even with GPU acceleration this decomposition is prohibitively
expensive; in the case of slender dumbbells and dilute suspensions,
we advocate a simple additive noise with a constant matrix $C$ as  a reasonable
approximation in the dilute limit as the off-diagonal terms in Eq.~\eqref{e:Mazur}
are negligible. We compared the full Cholesky decomposition
and an additive-noise approximation in various test runs and found that the results for 
the collective mean square displacement differed by only a few percent.

\section{Conclusions}

We have examined the stroke-averaged, far-field equations of motion for symmetric
dumbbells, and verified the general properties of this coarse-grained 
model by comparing with microscopic numerical simulations at relatively low
densities.  Remarkably, the microscopic  and coarse-grained simulations agree well even at intermediate-to-high swimmer densities, 
where the effective equations of motion are expected to become less accurate.  However, it should be kept
in mind that at very high densities, when collisions (i.e., steric
effects) become relevant, lubrication effects as well as near-field
hydrodynamics must be modelled more carefully.
\par
In the case of dumbbells arranged on a 3d grid, the translational speed
due to hydrodynamic interaction between dumbbells varies predictably
with spacing, tending toward $|D|^{-4}$ decay, where $|D|$ is the
distance between dumbbell centres.  Due the short range of the effective hydrodynamic
interactions for symmetric dumbbells, adding more swimmers at a fixed
density has only a minimal impact on dumbbell translational speed. On the other hand, the  
collective swimming speed can be noticeably increased by replacing a randomized phase 
distribution with an ordered, ``optimal'' distribution of
phases such that the difference in phase between a periodically-driven
dumbbell and its nearest neighbors is $\pi/2$. 
\par
Generally, our numerical investigations illustrate that GPU-based simulations 
of multi-swimmer systems can provide a valuable tool for studying 
collective motions at very low Reynolds number. Moreover, the CUDA algorithm  used in our computer experiments 
can be readily adapted to simulate hydrodynamic interactions between colloids  
that can be trapped and manipulated by means of optical tweezers~\cite{2009LeEtAl}. 
Such theoretical investigations can help to create more efficient micropumps, e.g., by 
optimizing the phase relations  in oscillating arrays of colloids.
\par
Finally, another long-term objective is to compare many-swimmer
simulations with predictions of effective field
theories~\cite{2002Ra}. Our above results suggest that the most
promising approach towards achieving this goal may be a two-step
procedure: (Step 1) One should try to derive stroke-averaged equations
of motion that correctly capture the phase dependence on the level of
effective two-particle interactions.  As our above discussion has
shown, such coarse-grained models can correctly reproduce many of the
main features of the microscopic model. Thus, it is sufficient
for many purposes to implement the coarse-grained equations into a
CUDA environment (step 2).  Compared to simulations of the full
microscopic dynamics, this may reduce the effective simulation time by
an additional factor of 100 or more since the analytic
stroke-averaging procedure makes it unnecessary to numerically resolve
the smallest dynamical time scales in the system.  We hope that our
analysis may provide useful guidance for future efforts in this
direction.

\paragraph*{Acknowledgements.--}
J. D. would like to thank Peter H\"anggi
for many stimulating discussions and the most enjoyable collaboration 
over the past years. This work was supported by the ONR, USA (J.D.).  V. P. acknowledges support from the
United States Air Force Institute of Technology.  The views expressed
in this paper are those of the authors and do not reflect the official
policy or position of the United States Air Force, Department of
Defense, or the U.S. Government. 

\appendix

\section{Stroke-averaging}

We derive the stroke-averaged effective interactions between two
symmetric, quasi-shape-driven dumbbell swimmers.  Each dumbbell
consists of two spheres of radius $a$. The swimming stroke of an
individual dumbbell is assumed to be both force-free and torque-free.

\subsection{One-dimensional case}
\label{a:SA-1d}
In one space dimension (1d) we denote the position of the spheres
belonging to dumbbell~$\gs$ by $X^\gs_s$, $s=1,2$.  To characterize
position and orientation of the dumbbell, we may introduce
center-of-mass and relative coordinates by 
\bse \be
R^\gs&=&\f{1}{2}\left(X_1^\gs+ X_2^\gs\right)
\\
S^\gs&=&X^\gs _2-X_1^\gs
\\
N^\gs&=&(X^\gs _2-X_1^\gs)/|X^\gs _2-X_1^\gs|; \ee hence, \be X_1^\gs=
R^\gs-S^\gs/2 \csp X_2^\gs= R^\gs+S^\gs/2 \ee which may also be
written as 
\be\label{e-a:1d_recovery} X^\gs_s= R^\gs+(-1)^s\; S^\gs/2.  
\ee 
\ese 
Furthermore, we define the vector connecting two swimmers $\gs$ and $\gr$ by 
\bse
\be D^{\gs\gr}&:=&R^\gs - R^\gr
\\
\hat{D}^{\gs\gr}&:=&(R^\gs - R^\gr)/|R^\gs - R^\gr|.  
\ee 
\ese 
The
force-free constraint for the dumbbell $\gs$ can be written as
\be\label{e:force_free} F_1^\gs=-F_2^\gs=:f^\gs, \ee with $F_s^\gs$
denoting the internal forces acting on the first and the second sphere
during a swimming stroke.  Neglecting thermal fluctuations, the 1d
equations of motion can be written as 
\be\label{a-e:eom_1d} \dot X^\gs_s = \sum_{\gr, r}H^{\gs\gr}_{sr} F^\gr_r.  \ee 
Here, we sum over
all swimmers $\gr=1,\ldots, S$ and the spheres $r=1,2$ of each
swimmer. Our goal is to derive from Eq.~\eqref{a-e:eom_1d} a
stroke-averaged effective equation of motion for $R^\gs$ (for
shape-driven dumbbells the motion of $S^\gs$ is trivial in 1d).
\par
The \lq\lq diagonal\rq\rq\space components of the hydrodynamic interaction tensor  $H$ are given by the inverse Stokes friction coefficient
\be
H^{\gs\gs}_{ss}= \gc^{-1}=(6\pi\mu a)^{-1}.
\ee
Adopting the Oseen approximation, the \lq\lq off-diagonal\rq\rq\space components ($s\ne r$) read 
\be
H^{\gs\gs}_{sr}=\f{\gk}{|S^\gs|}
\csp
H^{\gs\gr}_{sr}=\f{\gk}{|X_s^\gs-X_r^\gr|}
\ee
where $\gk=(4\pi\mu)^{-1}$. It is useful to rewrite 
\bse
\be
X_s^\gs- X_r^\gr
=
D^{\gs\gr}
+
Y^{\gs\gr}_{sr}
\ee
where 
\be
Y^{\gs\gr}_{sr}
:=
\f{1}{2}[(-1)^s\; S^\gs-(-1)^r\; S^\gr].
\ee
\ese
Using the force free condition~\eqref{e:force_free}, we obtain from Eq.~\eqref{a-e:eom_1d}
\bse
\be
\dot{R}^\gs
&=&
\sum_{\gr} 
\f{1}{2}[ (H^{\gs\gr}_{11} -H^{\gs\gr}_{12} )+ (H^{\gs\gr}_{21}-H^{\gs\gr}_{22})] f^\gr
\notag\\
&=:& 
\sum_{\gr} 
A^{\gs\gr} f^\gr
\label{e:1D_com}
\ee
and
\be
\dot{S}^\gs
&=&
\sum_{\gr} 
[(H^{\gs\gr}_{21}-H^{\gs\gr}_{22})-
(H^{\gs\gr}_{11} -H^{\gs\gr}_{12} )] f^\gr
\notag\\
&=:& 
\sum_{\gr} 
B^{\gs\gr} f^\gr
\label{e:1D_rel}
\ee
\ese
Considering approximately shape-driven dumbbells, we have
\be\label{e:shape_driven}
S^\gs&=&L^\gs(t)\,N^\gs, \notag \\
|S^\gs|&=&L^\gs(t), \notag \\
\dot{S}^\gs&=&\dot{L}^\gs(t)\,N^\gs,
\ee
where the periodic function  $L^\gs(t)>0$ describes the shape (length) of the dumbbell at time $t$. 
Hence, inverting~\eqref{e:1D_rel} we obtain the force as a function of the shape
\be\label{a-e:eom_f_1d}
f^\gr=\sum_\nu(B^{-1})^{\gr\nu}\dot{L}^\nu(t)\,N^\nu,
\ee
where $B^{-1}$ denotes the inverse of the
$(S\times S)$-matrix $B:=(B^{\gs\gr})$ defined in~\eqref{e:1D_rel}. 
Substituting this result into Eq.~\eqref{e:1D_com} yields the 
following closed equations for the position coordinates
\be\label{a-e:eom_R_1d}
\dot{R}^\gs=
\sum_{\gr,\nu}A^{\gs\gr} (B^{-1})^{\gr\nu}\dot{L}^\nu(t)\,N^\nu.
\ee
By means of Eqs.~\eqref{e:shape_driven}, we can rewrite the off-diagonal components of the Oseen tensor as
\bse
\be
H^{\gs\gs}_{sr}=\f{\gk}{L^\gs},
\qquad
H^{\gs\gr}_{sr}=\f{\gk}
{| D^{\gs\gr}+Y^{\gs\gr}_{sr}|}
\ee
where
\be\label{e-a:1d_split}
Y^{\gs\gr}_{sr}
=
\f{1}{2}[(-1)^s\; L^\gs(t)\,N^\gs-(-1)^r\; L^\gr(t)\,N^\gr] \notag
\\
\ee
\ese
For a system consisting of more than two dumbbells ($S>2$), the rhs. of 
Eq.~\eqref{a-e:eom_f_1d} contains not only two-body, but also three-body, 
four-body, \dots, $S$-body contributions. However, focussing only on the 
dominant two-body contributions,  $B:=(B^{\gs\gr})$ can be exactly inverted 
and the rhs. of Eq.~\eqref{a-e:eom_R_1d} can be expanded in the low-density 
limit corresponding to  $|D^{\gs\gr}|\to\infty$. Averaging the resulting power 
series over a stroke period  $[t-T/2,t+T/2]$ as described in Sec.~\ref{s:SA} 
and keeping only the leading order contribution, we find the following 
1d stroke-averaged equation of motion in two-body approximation
$$
\dot{R}^\gs
\simeq
\f{9}{16} a\go \sum_{\gr}
\sin(\gvf^\gs-\gvf^\gr)
\left(\f{\gl}{\ell}\right)^2  
\biggl(\f{\ell}{|D^{\gs\gr}|}\biggr)^4 \hat{D}^{\gs\gr}.
$$

\subsection{Three-dimensional case}
\label{a:3d}

In the 3d case, the derivation of stroke-averaged equations becomes more complicated due to the
additional rotational degrees of freedom. 
\par
As before, we consider a dilute suspension of $\gs=1,\ldots, N$ geometrically identical  dumbbells of 
prescribed length $ L^\gs(t)$. To characterize the motion of the  dumbbells, we define position and orientation vectors by
\be\label{e-a:3d_coordinates}
\bs R^\gs(t)&:=&\f{1}{2}(\bs X^{\gs1} + \bs X^{\gs2}) \notag
\\
\bs N^\gs(t)&:=&\f{\bs S^\gs}{|\bs S^\gs|},
\ee
with $\bs S^\gs$ denoting the  non-normalized orientation vector, i.e., for a shape-driven dumbbell we can write
\bse
\be
\bs S^\gs(t)&:=&\bs X^{\gs 2}-\bs X^{\gs1}= L^\gs(t)\bs N^\gs,\\
S^\gs(t)&:=&|\bs S^\gs|= L^\gs.
\ee
Similar to Eq.~\eqref{e-a:1d_recovery}, we can recover the bead coordinates $\{\bs X^{\gs 1},\bs X^{\gs 2}\}$ from  $\{\bs R^\gs,\bs N^\gs\}$ by means of
\be\label{e-a:3d_cord_inv}
\bs X^{\gs s}&=&\bs R^\gs +(-1)^s \bs N^\gs  L^\gs/2.
\ee
\ese
\par
As before, we consider shape-driven  dumbbells with \mbox{$ L^\gs(t)=\ell+\gl\sin(\go t+\gvf^\gs)$}.  
From the definition~\eqref{e-a:3d_coordinates}, one then finds that the exact 
equations of motion for $\{\bs R^\gs,\bs N^\gs\}$ are given by
\bse\label{e-a:eom_3d_exact}
\be
\dot{R}^\gs_i
&=&
\f{1}{2}\sum_{s,\gr,r} H_{ij}^{(\gs s)(\gr r)}F^{\gr r}_j\\
\dot{N}_i^\gs
&=&\notag
(\gd_{ik}- N^\gs_iN^\gs_k)
\times
\\
&&
\f{1}{ L^\gs}
\sum_{\gr\ne\gs, r}\left[
H_{kj}^{(\gs 2)(\gr r)} - H_{kj}^{(\gs 1)(\gr r)} \right] F^{\gr r}_j \notag
\\
\ee
\ese
The indices $s,r\in\{1,2\}$ label the spheres and, throughout,  we use the sum 
convention $H_{ij} F_j:=\sum_j H_{ij}F_j$ for spatial tensor indices.  Restricting 
ourselves to dilute suspensions of slender  dumbbells, we adopt  the Oseen 
approximation for the hydrodynamic interaction tensor, i.e.,
\bse\label{e:3D_oseen}
\be
H^{(\gs s)(\gs s)}_{ij}&=&
(6\pi\mu a)^{-1}\gd_{ij},
\\
H^{(\gs s)(\gs r)}_{ij}&=&\f{\gk}{ L^\gs}\left(\gd_{ij}+N^\gs_i N^\gs_j \right),
\\
H^{(\gs s)(\gr r)}_{ij}&=&\f{\gk}{|\bs X^{\gs s}-\bs X^{\gr r} |}
\times\notag
\\&&
\left[\gd_{ij}+\f{ (X_i^{\gs s}-X_i^{\gr r}) (X_j^{\gs s}-X_j^{\gr r})}{|\bs X^{\gs s}-\bs X^{\gr r} |^2} \right] \notag
\\
\label{e:Oseen_c}
\ee 
\ese
where $\gk=(8\pi\mu)^{-1}$. To obtain from Eqs.~\eqref{e-a:eom_3d_exact} closed stroke-averaged equations for $\left\{\bs R^\gs,\bs N^\gs\right\}$, we must 
\begin{itemize}
\item[$a.$] perform a far-field expansion of the hydrodynamic interaction tensor;
\item[$b.$] determine the internal forces $\bs F^{\gs s}$, required to maintain the  dumbbells' prescribed shape $ L^\gs(t)$;
\item[$c.$] expand the resulting equations in powers of $(\gl/\ell)$ and average over a  stroke period $[t,t+T].$
\end{itemize}

\subsubsection{Far-field expansion}

The Oseen tensor components $H_{ij}$ given in Eq.~\eqref{e:3D_oseen} are functions of the sphere separation vectors $\bs X^{\gs s}-\bs X^{\gr r}$. By means of Eq.~\eqref{e-a:3d_cord_inv}, we may decompose
\be
\bs X^{\gs s}-\bs X^{\gr r}
=
\bs D^{\gs\gr}+\bs Y^{(\gs s)(\gr r)},
\ee
where similar to Eqs.~\eqref{e-a:1d_split} we have defined
\bse
\be
\bs D^{\gs\gr}
&:=& \bs R^{\gs }-\bs R^{\gr},
\\
\bs Y^{(\gs s)(\gr r)}&:=& \f{1}{2}\left[
(-1)^s \bs N^\gs  L^\gs   - (-1)^r \bs N^\gr  L^\gr 
\right] \notag
\\
\ee
\ese
Then, for $\gs\ne \gr$, the Oseen tensor components~\eqref{e:Oseen_c} take the form
\be
H_{ij}
:=
\f{\gk}{|\bs D+\bs Y |} 
\biggl(
\gd_{ij} +
\f{D_i + Y_i}{|\bs D+\bs Y|} 
\f{D_j +  Y_j}{|\bs D +\bs Y|} 
\biggr).
\qquad
\ee
For clarity, we dropped superscripts here using the abbreviations $\bs Y:=\bs Y^{(\gs s)(\gr r)}$ 
and $\bs D:=\bs D^{\gs\gr}$. In the dilute limit, corresponding to $|\bs Y|\ll |\bs D|$ we 
may perform a far-field (Taylor) expansion of the tensor components $H_{ij}$. For this purpose we define
\bse
\be
H^0_{ij}:=H_{ij}(\bs Y=\bs 0)
=
\f{\gk}{|\bs D|} 
\left(\gd_{ij}+\hat{D}_i \hat{D}_j\right),
\qquad
\ee
where 
\be
\hat{D}_i
:=
\f{D_i}{|\bs D|}
\ee
\ese
is the unit vector in the direction of $\bs D$. Reinstating upper indices, the formal Taylor expansion 
of $H^{(\gs s)(\gr r)}_{ij}$ at \mbox{$\bs Y=\bs 0$}  can be expressed as
\bse\label{e:expansion}
\be
H^{(\gs s)(\gr r)}_{ij}
=
\sum_{q=0}^\infty H^{\gs\gr}_{ij,k_q... k_1} 
 Y^{(\gs s)(\gr r)}_{k_1} \cdots Y^{(\gs s)(\gr r)}_{k_q},\notag
\\
\ee
where
\be
H^{\gs\gr}_{ij,k_q... k_1}
&:=&
\f{1}{q!}\,
\p_{k_1}\cdots \p_{k_q} H^0_{ij}\biggl|_{\bs D=\bs D^{\gs\gr}}
\ee
\ese
and $\p_{k_j}:=\p/\p D_{k_j}$. Explicit expressions for the expansion coefficients $H_{ij,k_q... k_1}$ with $q=1,2,3,4$ are summarized in ~\ref{a:Oseen_derivative}.  The expansion~\eqref{e:expansion}  will be used in the next part to compute the interaction forces  $\bs F^{\gs s}$, and, later on, it will also be inserted into the exact equations of motion~\eqref{e-a:eom_3d_exact}.

\subsubsection{Internal forces in two-particle approximation}

We wish to determine the internal forces $\bs F^{\gs s}$ in Eq.~\eqref{e-a:eom_3d_exact} by means of an iterative procedure, restricting ourselves to two-body interactions and assuming, as usual, that individual  dumbbell swimmers are both force-free and torque-free, i.e.,
\bse
\be
0&\overset{!}{\equiv}&\label{e:force-free}
\sum_s F_i^{\gs s},\\
0&\overset{!}{\equiv}&\label{e:torque-free}
T^\gs_i(\bs y)
:=
\sum_{s}\eps_{ijk} (X^{\gs s}_j-y_j)F^{\gs s}_k,
\qquad
\ee
\ese
where $\bs y=(y_j)$ is an arbitrary reference point.
Substituting Eq.~\eqref{e:force-free} into  Eq.~\eqref{e:torque-free} we find that
\be\notag
0\equiv\eps_{ijk} (X^{\gs 1}_j-X^{\gs 2}_j) F^{\gs 1}_k,
\ee
or equivalently
\be
0\equiv\eps_{ijk} N^{\gs}_j F^{\gs 1}_k.
\ee
This implies that  $\bs F^{\gs s}$ must be of the form
\be\label{e-a:force_ansatz_db}
\bs F^{\gs s}= f^{\gs s}\bs N^\gs,
\qquad
\label{e-a:force_free_db}
f^{\gs 2}=-f^{\gs 1}.
\ee
It thus remains to express the $N$ unknown functions $f^{\gs 1}$ in terms of $\left\{\bs R^\gs,\bs N^\gs\right\}$.

\paragraph*{Shape-constraints.--}
To determine the unknown functions $f^{\gs 1}$, we exploit the $N$ independent shape constraints
\be\label{e:constraint_db}
 \dot{L}^\gs
\overset{!}{=}
N^\gs_i(\dot X^{\gs 2}_i-\dot X^{\gs 1}_i).
\ee
Inserting the equations of motion for $\bs X^{\gs s}$, we find the explicit condition
\be\label{e:constraint_db_a}
 \dot{L}^\gs
\overset{!}{=}
\sum_{\gr,r}
N^\gs_i
\left[ H_{ij}^{(\gs 2)(\gr r)}-H_{ij}^{(\gs 1)(\gr r)}  \right]N^\gr_j\,f^{\gr r}.
\ee
Introducing the convenient abbreviation
\be
h^{(\gs s)(\gr r)}:=
N^\gs_i\;
H_{ij}^{(\gs s)(\gr r)}
\,
N^\gr_j,
\ee
we can write Eq.~\eqref{e:constraint_db_a} as
\be
 \dot{L}^\gs
&\overset{!}{=}&
\sum_{r}
\left[
h^{(\gs2)(\gs r)}-h^{(\gs1)(\gs r)}
\right]
f^{\gs r}+
\notag\\
&&
\sum_{\gr\ne \gs,r}
\left[
h^{(\gs2)(\gr r)}-h^{(\gs1)(\gr r)}
\right]
f^{\gr r}.
\qquad
\label{e:constraint_b}
\ee
Here, we have separated interactions within the dumbbell~$\gs$ from those with other swimmers $\gr\ne\gs$. Using the force-free constraint~\eqref{e-a:force_free_db}, Eq.~\eqref{e:constraint_b} takes the form
\bse\label{e:constraint_c}
\be\label{e:constraint_c_1}
 \dot{L}^\gs
&\overset{!}{=}&
b^{\gs\gs} f^{\gs 1} + \sum_{\gr\ne \gs}
b^{\gs\gr}
f^{\gr 1},
\ee
with coefficient functions
\be\label{e:b-coefficients}
b^{\gs\gr}
:=\notag
h^{(\gs2)(\gr 1)}+
h^{(\gs1)(\gr 2)}
-
[h^{(\gs2)(\gr 2)}
+h^{(\gs1)(\gr 1)}] \notag
\\
\ee
\ese
The $N$ linear equations~\eqref{e:constraint_c_1} determine the $N$ unknown functions $f^{\gr 1}$ by means of an iterative procedure.

\paragraph*{Iteration scheme.--} Rewriting Eq.~\eqref{e:constraint_c_1} in the form
\be
 f^{\gs 1}= 
\f{ \dot{L}^\gs} {b^{\gs\gs}} - \sum_{\gr\ne \gs} \f{b^{\gs\gr}}{b^{\gs\gs}} f^{\gr 1}
\ee
we obtain the following recursive sequence
\be
 f^{\gs 1}_{(n)}= 
\f{ \dot{L}^\gs} {b^{\gs\gs}} - \sum_{\gr\ne \gs} \f{b^{\gs\gr}}{b^{\gs\gs}} f^{\gr 1}_{(n-1)}.
\ee
Starting from the initial condition $f^{\gr 1}_{(0)}=0$, the first iteration gives the force generated by an isolated, shape-driven dumbbell  
\be
f^{\gs 1}_{(1)}= 
\f{ \dot{L}^\gs} {b^{\gs\gs}}. 
\ee
The second iteration yields a correction due to pair interactions with other dumbbells,
\be
f^{\gs 1}_{(2)}
&=& \notag
\f{ \dot{L}^\gs} {b^{\gs\gs}}  - \sum_{\gr\ne \gs} \f{b^{\gs\gr}}{b^{\gs\gs}} \f{ \dot{L}^\gr} {b^{\gr\gr}} 
\\
&=& 
f^{\gs 1}_{(1)} - \sum_{\gr\ne \gs} \f{b^{\gs\gr}}{b^{\gs\gs}} \f{ \dot{L}^\gr} {b^{\gr\gr}}.
 \label{e:2nd_iteration}
\ee
Similarly, one obtains from the third iteration 
\be
f^{\gs 1}_{(3)}
&=&\notag 
\f{ \dot{L}^\gs} {b^{\gs\gs}} - \sum_{\gr\ne \gs} \f{b^{\gs\gr}}{b^{\gs\gs}}
\left[ 
\f{ \dot{L}^\gr} {b^{\gr\gr}}  - \sum_{\nu\ne \gr} \f{b^{\gr\nu}}{b^{\gr\gr}} \f{\dot{\ell}^\nu} {b^{\nu\nu}} 
\right]
\\
&=& 
f^{\gs 1}_{(2)}+ \sum_{\gr\ne \gs}  \sum_{\nu\ne \gr} 
 \f{b^{\gs\gr}}{b^{\gs\gs}} \f{b^{\gr\nu}}{b^{\gr\gr}} \f{\dot{\ell}^\nu} {b^{\nu\nu}}. 
 \label{e:3rd_iteration}
\ee
The last term can be interpreted as a three-particle interaction correction.  Let us assume that the system contains $\gs=1,\ldots,N$ dumbbells. Then, as evident from the  \lq exclusive\rq~ summation in Eq.~\eqref{e:3rd_iteration}, the iteration will approach a fixed point after $N$ iterations,
\be
f^{\gs1}_{(N+1)}=f^{\gs 1}_{(N)}.
\ee
The fixed point $f_{(N)}$ corresponds to the exact solution, i.e.,  $f_{(N)}$ is the internal force generated by a  dumbbell in order to maintain its prescribed shape in the presence hydrodynamic forces of $N-1$ other  dumbbells. In the remainder, we shall restrict ourselves to considering one-particle and two-particle interactions, corresponding to~$f^{\gs1}_{(1)}$ and~$f^{\gs1}_{(2)}$.

\paragraph*{Coefficients $b^{\gs\gr}$.--}
We still need to determine the coefficients $b^{\gs\gr}$ from~\eqref {e:b-coefficients}.  The 'diagonal' coefficients $b^{\gs\gs}$ can be calculated exactly by
noting that
\bse\label{e:h-symm}
\be
h^{(\gs s)(\gs s)}
&=&(6\pi\mu a)^{-1}
=\f{4\gk}{3 a}
\\
h^{(\gs 1)(\gs 2)} &=& h^{(\gs 2)(\gs 1)}
\quad=
\f{2\gk}{ L^\gs} 
\ee
\ese
We thus have
\be
b^{\gs\gs}=\f{4\gk}{ L^\gs}\left( 1-\f{2 L^\gs}{3a}\right).
\ee
In order to determine the coefficients $b^{\gs\gr}$ with $\gr\ne\gs$, we need to use the far-field expansion~\eqref{e:expansion}. Defining the contraction
\be\label{e:hq}
h^{\gs\gr }_{k_q... k_1}&:=&
N^\gs_i N^\gr_j\; H^{\gs\gr}_{ij,k_q... k_1}
\ee
allows us to write
\bse
\be
b^{\gs\gr}
=
\sum_{q=0}^\infty 
b_q^{\gs\gr},
\ee
where
\be
&&b_q^{\gs\gr}
=\notag
h^{\gs\gr }_{k_q... k_1} \times
\\
&&\notag\quad
\bigl\{
Y^{(\gs 1)(\gr 2)}_{k_1} \cdots Y^{(\gs 1)(\gr 2)}_{k_q} +
Y^{(\gs 2)(\gr 1)}_{k_1} \cdots Y^{(\gs 2)(\gr 1)}_{k_q} -
\\
&&\quad\notag\;
[Y^{(\gs 1)(\gr 1)}_{k_1} \cdots Y^{(\gs 1)(\gr 1)}_{k_q} +
Y^{(\gs 2)(\gr 2)}_{k_1} \cdots Y^{(\gs 2)(\gr 2)}_{k_q}]
\bigr\}.
\\
\label{e:bq}
\ee
\ese
We define
\bse
\be\label{e:N2}
N^{\gs\gr}_{k_1k_2} &:=&
N^\gs_{k_1} N^\gr_{k_2}+N^\gs_{k_2} N^\gr_{k_1}\;,
\\
N^{\gs\gs\gr}_{k_1k_2 k_3} &:=&\notag
N^\gs_{k_1}N^\gs_{k_2}N^\gr_{k_3}+
N^\gs_{k_1}N^\gs_{k_3}N^\gr_{k_2}+
\\
&&
N^\gs_{k_2}N^\gs_{k_3}N^\gr_{k_1},
\\
N^{\gs\gs\gr\gr}_{k_1k_2 k_3 k_4} &:=&\notag
N^\gs_{k_1}N^\gs_{k_2}N^\gr_{k_3} N^\gr_{k_4}+
N^\gs_{k_1}N^\gs_{k_3}N^\gr_{k_2} N^\gr_{k_4}+
\\
&&\notag
N^\gs_{k_1}N^\gs_{k_4}N^\gr_{k_2} N^\gr_{k_3}+
N^\gs_{k_2}N^\gs_{k_3}N^\gr_{k_1} N^\gr_{k_4}+
\\
&&\notag
N^\gs_{k_2}N^\gs_{k_4}N^\gr_{k_1} N^\gr_{k_3}+
N^\gs_{k_3}N^\gs_{k_4}N^\gr_{k_1} N^\gr_{k_2},
\\
\\
N^{\gs\gs\gs\gr}_{k_1k_2 k_3 k_4} &:=&\notag
N^\gs_{k_1}N^\gs_{k_2}N^\gs_{k_3} N^\gr_{k_4}+
N^\gs_{k_1}N^\gs_{k_2}N^\gs_{k_4} N^\gr_{k_3}+
\\
&& \notag
N^\gs_{k_1}N^\gs_{k_3}N^\gs_{k_4} N^\gr_{k_2}+
N^\gs_{k_2}N^\gs_{k_3}N^\gs_{k_4} N^\gr_{k_1}.
\\
\ee
\ese
With these abbreviations we find
\bse
\be
b^{\gs\gr}_0&=&0,
\\
b^{\gs\gr}_1&=&0,
\\
b^{\gs\gr}_2
&=&
h^{\gs\gr }_{k_2 k_1}\; L^\gs  L^\gr\; N^{\gs\gr}_{k_1k_2},
\\
b^{\gs\gr}_3 &=&0,
\\
b^{\gs\gr}_4
&=&\notag
\f{1}{4}h^{\gs\gr }_{k_4k_3k_2k_1}
\bigl[L^\gs L^\gs L^\gs  L^\gr \,
N^{\gs\gs\gs\gr}_{k_1k_2 k_3 k_4} +
\\&&\qquad\qquad\quad \notag
L^\gr  L^\gr L^\gr L^\gs\,
N^{\gr\gr\gr\gs}_{k_1k_2 k_3 k_4} 
\bigr],
\\
\ee
\ese
which can be used in~\eqref{e:2nd_iteration}.

\subsubsection{Stroke-averaging}

\paragraph*{Translational motion.--}
Inserting the ansatz~\eqref{e-a:force_ansatz_db} into Eq.~\eqref{e-a:eom_3d_exact}, the motion of the position coordinate is determined by 
\be
\dot{R}^\gs_i
=
\f{1}{2}\sum_{s,\gr,r}H_{ij}^{(\gs s)(\gr r)}f^{\gr r} N^\gr_j.
\ee
It is convenient to consider \lq internal\rq\space and external contributions separately by writing
\bse
\be
\label{e:dot_R}
\dot{R}^\gs_i&=&
N^\gs_i  I^\gs
 +
\sum_{\gr\ne\gs}J^{\gs\gr}_i
\ee
where
\be
I^\gs
&:=& 
\f{1}{2}\sum_{s,r} h^{(\gs s)(\gs r)}f^{\gs r},
\\
J^{\gs\gr}_i
&:=& 
\f{1}{2}\sum_{s,r} H_{ij}^{(\gs s)(\gr r)}f^{\gr r} N^\gr_j.
\label{e:def_J}
\ee
\ese
Here we have used that
\be\notag
H_{ij}^{(\gs s)(\gs r)}N^\gs_j
=h^{(\gs s)(\gs r)}N^\gs_i.
\ee
Using the force-free constraint and Eq.~\eqref{e:h-symm}, we find
\be
I^\gs =0,
\ee
i.e., the only contribution to the translation of swimmer $\gs$ comes from interactions with the other dumbbells $\gr\ne\gs$.
Hence, we still need to determine the second contribution $J^{\gs\gr}_i$ from Eq.~\eqref{e:def_J}, which can be written in the form
\bse
\be
J^{\gs\gr}_i
=
\Xi^{\gs\gr}_{ij}\,
N^\gr_j\; f^{\gr 1},
\label{e:def_J_2}
\ee
where
\be
\Xi^{\gs\gr}_{ij}
&:=&
\f{1}{2}\bigl[H_{ij}^{(\gs 1)(\gr 1)}-H_{ij}^{(\gs 1)(\gr 2)}\bigr]+
\notag\\&& 
\f{1}{2}\bigl[H_{ij}^{(\gs 2)(\gr 1)}-H_{ij}^{(\gs 2)(\gr 2)}\bigr].
\ee
\ese
Inserting the far-field expansion for the Oseen tensor, we obtain
\bse
\be
\Xi^{\gs\gr}_{ij}
=
\sum_{q=0}^\infty H^{\gs\gr}_{ij,k_q\ldots k_1} \;P^{\gs\gr}_{k_1\ldots k_q}(\bs Y)
\ee
with polynomials $P^{\gs\gr}_{k_1\ldots k_q}$ given by
\be
P^{\gs\gr}_{k_1\ldots k_q}(\bs Y)
&:=&\notag
\f{1}{2} Y^{(\gs 1)(\gr 1)}_{k_1} \cdots Y^{(\gs 1)(\gr 1)}_{k_q} -
\\&&\notag
\f{1}{2} Y^{(\gs 1)(\gr 2)}_{k_1} \cdots Y^{(\gs 1)(\gr 2)}_{k_q} +
\\&&\notag
\f{1}{2}  Y^{(\gs 2)(\gr 1)}_{k_1} \cdots Y^{(\gs 2)(\gr 1)}_{k_q} -
\\&&
\f{1}{2} Y^{(\gs 2)(\gr 2)}_{k_1} \cdots Y^{(\gs 2)(\gr 2)}_{k_q} .
\qquad\quad
\ee
\ese
In particular, for $q=0$ we have $P^{\gs\gr}=0$ and for $q\ge 1$
\bse
\be
P^{\gs\gr}_{k_1}&=& L^\gr N^\gr_{k_1},
\\
P^{\gs\gr}_{k_1k_2}&=&0,
\\
P^{\gs\gr}_{k_1k_2 k_3}&=&\notag
\f{1}{4} \bigl[ L^\gs  L^\gs  L^\gr\; N^{\gs\gs\gr}_{k_1k_2 k_3} +
\\&&\quad 
L^\gr L^\gr L^\gr\; N^\gr_{k_1}N^\gr_{k_2}N^\gr_{k_3} \bigr],
\quad
\\
P^{\gs\gr}_{k_1\ldots k_4}&=&0.
\ee
\ese
Since Eq.~\eqref{e:dot_R} already contains a sum over $\gr$, neglecting three-body effects means that, in order to compute~$J^{\gs\gr}_i$, we should use $f^{\gr 1}\simeq f^{\gr 1}_{(1)}= \dot{L}^\gr/b^{\gr\gr}$ in Eq.~\eqref{e:def_J_2}. After averaging~\eqref{e:def_J_2} over period, we obtain at leading order of $(\ell/|D|)$
\be\label{e:J_3}
\ovl{J^{\gs \gr}_i}\simeq
\f{1}{4}\,
N^\gr_j\; H^{\gs\gr}_{ij,k_3k_2k_1} 
N^{\gs\gr}_{k_1k_2 k_3} \;\ovl{\f{( L^\gs)^2   L^\gr  \dot{L}^\gr}{b^{\gr\gr}} },
\quad
\ee
where to leading order in $\gl$
\bse\label{a-e:J_contributions}
\be\label{a-e:J_contributions-a}
\ovl{\f{( L^\gs)^2   L^\gr  \dot{L}^\gr}{b^{\gr\gr}} }
\simeq
-\go a \f{3\ell^2\gl^2}{8\gk} \sin(\gvf^\gs-\gvf^\gr).
\ee
The contraction is obtained as
\be
&\notag
 \f{1}{4}N^\gr_j\;H^{\gs\gr}_{knl} N^{\gs\gr}_{lnk}
=
-
\f{3\gk}{8 |\bs D|^4}
\bigl\{
N^\gs_{i}
(2s+4qr-10 sr^2) +
\\
& \notag
\hat D_i(1+ 2 q^2-5 s^2-5r^2- 20qsr  + 35 s^2 r^2)
\bigr\}.
\\
\ee
where $\bs D:=\bs D^{\gs\gr}:=\bs R^\gs-\bs R^\gr$, $\hat{\bs D}:={\bs D^{\gs\gr}}/{|\bs D^{\gs\gr}|}$
and
\be\notag
s=\hat{D}^{\gs\gr}_jN^\gs_j,
\qquad
r=\hat{D}^{\gs\gr}_jN^\gr_j,
\qquad
q=N^\gs_jN^\gr_j
\ee
\ese
denote the three possible pairwise projections of the relevant unit vectors $\bs N^\gs$, $\bs N^\gr$, and $\hat{\bs D}^{\gs\gr}$. Inserting Eqs.~\eqref{a-e:J_contributions} into \eqref{e:J_3} yields the expression for $\ovl{J^{\gs \gr}_i}$ that is given in Eq.~\eqref{J}.

\paragraph*{Change of orientation.--}
The exact equations of motion for the orientation vectors $\bs N^\gs$ read
\bse
\be\label{a-e:dot_N}
\dot{N}_i^\gs
=
(\gd_{ik}- N^\gs_iN^\gs_k)
\sum_{\gr\ne\gs} G^{\gs \gr}_k,
\ee
where
\be
G^{\gs \gr}_k
:=
N^\gr_j 
\sum_b\left[
H_{kj}^{(\gs 2)(\gr b)} -H_{kj}^{(\gs 1)(\gr b)} \right] \f{f^{\gr b}}{ L^\gs}.
\ee
\ese
Using the force-free constraint~\eqref{e:force-free}, one obtains explicitly
\be
G^{\gs \gr}_k
&=&\notag
N^\gr_j 
\bigl[
-H_{kj}^{(\gs 1)(\gr 1)} +H_{kj}^{(\gs 1)(\gr 2)} +
\\&&
\qquad\quad
H_{kj}^{(\gs 2)(\gr 1)} -H_{kj}^{(\gs 2)(\gr 2)} \bigr] \f{f^{\gr 1}}{ L^\gs}.
\ee
Inserting the expansion for hydrodynamic tensor gives
\be
&&\notag
-H_{ij}^{(\gs 1)(\gr 1)}
+Hyd_{ij}^{(\gs 1)(\gr 2)}
+H_{ij}^{(\gs 2)(\gr 1)}
-H_{ij}^{(\gs 2)(\gr 2)} 
\\\notag
&=&
\sum_{q=0}^\infty H^{\gs\gr}_{ij,k_q... k_1} \times
\\\notag
&&
\bigl[
- Y^{(\gs 1)(\gr 1)}_{k_1} \cdots Y^{(\gs 1)(\gr 1)}_{k_q} +
 Y^{(\gs 1)(\gr 2)}_{k_1} \cdots Y^{(\gs 1)(\gr 2)}_{k_q} +
 \\&&\;
 Y^{(\gs 2)(\gr 1)}_{k_1} \cdots Y^{(\gs 2)(\gr 1)}_{k_q} -
 Y^{(\gs 2)(\gr 2)}_{k_1} \cdots Y^{(\gs 2)(\gr 2)}_{k_q} 
 \bigr].\notag
 \qquad
\ee
The polynomial terms in brackets are exactly those encountered earlier in Eq.~\eqref{e:bq}.  Hence, the first two non-vanishing contributions come from $q=2$ and $q=4$. Neglecting three-body effects  means that, similar to above, we should use $f^{\gr 1}\simeq f^{\gr 1}_{(1)}= \dot{L}^\gr/b^{\gr\gr}$. Hence, truncating after $q=4$ we have
\be
G^{\gs \gr}_i
&\simeq&\notag
N^\gr_j\,
H^{\gs\gr}_{ij,k_2k_1}\,
N^{\gs\gr}_{k_1 k_2} 
\f{ L^\gr \dot{L}^\gr}{b^{\gr\gr}}
+ 
N^\gr_j\;H^{\gs\gr}_{ij,k_4 k_3k_2k_1}  
\times
\\
&&
\f{1}{4}\left[
N^{\gs\gr}_{k_1k_2 k_3 k_4} \;( L^\gs)^2   L^\gr
+
N^{\gr\gs}_{k_1k_2 k_3 k_4}\;( L^\gr)^3
\right]
\f{ \dot{L}^\gr}{b^{\gr\gr}}.
\qquad\notag
\ee
Averaging this expression over a period, we find 
\be
\ovl{G^{\gs \gr}_i}=\f{1}{4}\,
N^\gr_j\;H^{\gs\gr}_{ij,k_4 k_3k_2k_1} 
N^{\gs\gr}_{k_1k_2 k_3 k_4} \;\ovl{\f{( L^\gs)^2   L^\gr  \dot{L}^\gr}{b^{\gr\gr}} }.
\qquad
\ee
The time average on the rhs. is the same as in~\eqref{a-e:J_contributions-a}. 
Exploiting the symmetry of lower indices of $N^{\gs\gr}_{mlnk}$, we obtain
\be
\lefteqn{\f{1}{4}N^\gr_j H_{ij,knlm} N^{\gs\gr}_{mlnk}
=\notag
\f{\gk}{4|\bs D|^5}\times}
\\\notag
& & \biggl\{
 N^\gs_i(-3+ 6q^2 +15 s^2+15r^2
\\\notag
& & \qquad -105 s^2r^2+60srq) +
\\\notag
& & 5\hat D_i \bigl(  3s +6rq+ 6sq^2   
-7s^3 -21sr^2
\\\notag
& & \qquad -42qs^2r + 63 s^3r^2
\bigr)
\biggr\}
\ee
Contracting with the orthogonal projector $(\gd_{ki}-N^\gs_k N_i^\gs)$, see Eq.~\eqref{a-e:dot_N}, eliminates 
terms proportional to  $N^\gs_i$, thus yielding  Eq.~\eqref{e:eom_b}.

\section{Partial derivatives of the Oseen tensor}
\label{a:Oseen_derivative}

This part summarizes the partial derivatives of the Oseen tensor that are required in the derivation of the far-field, stroke-averaged equations of motion~\eqref{e:eom}, see Eq.~\eqref{e:expansion} in ~\ref{a:3d}.
\par
Consider the distance vector $\bs D=(D_k)$, its  associated unit vector $(\hat{D}_k)$ and orthogonal projector $(\Pi_{ik})$, given by
\be
\hat{D}_k:=\f{D_k}{|\bs D|}
,\qquad
\Pi_{ik}:=\gd_{ik}-\hat{D}_i\hat{D}_k.
\ee
We wish to compute the partial derivatives of the Oseen tensor
\be
H_{ij}:=\f{\gk}{|\bs D|}\left(1+ \hat{D}_i \hat{D}_j\right)
\ee
where $\gk:=(8\pi\mu)^{-1}$. Abbreviating $\p_k:=\p/\p D_k$, we have
\bse
\be
\p_k |\bs D|&=&\f{D_k}{|\bs D|}=\hat{D}_k
\\
\p_k\hat{D}_i&=& \f{\gd_{ik}}{|\bs D|} -\f{D_kD_i}{|\bs D|^3}=\f{\Pi_{ik}}{|\bs D|}
\\
\p_n \Pi_{ik}&=&
-\f{1}{|\bs D|}\left(\hat{D}_i \Pi_{nk}+\hat{D}_k \Pi_{ni}\right)
\ee
\ese

\paragraph*{First order derivatives.--}
A straightforward calculation gives
\be
H_{ij,k}
&:=&\notag
\p_kH_{ij}
\\&=&\notag
-\f{\hat{D}_k}{|\bs D|}H_{ij}
+
\f{\gk}{|\bs D|^2}\left(\Pi_{ik}\hat D_j+\Pi_{jk}\hat D_i\right)\\
&=&\notag
\f{\gk}{|\bs D|^2}
\bigl(-\hat{D}_k\gd_{ij}+\hat D_j\gd_{ik}+\hat D_i\gd_{jk}
\\&&
-3\times\hat{D}_k\hat{D}_i \hat{D}_j\bigr).
\ee

\paragraph*{Second order derivatives.--}
The second order derivatives, normalized by $n!$ with $n=2$, are defined by
\be
H_{ij,kn}
&:=&\notag
\f{1}{2!}\p_n\p_kH_{ij}
\ee
and read explicitly
\be
H_{ij,kn}
&=&\notag
\f{\gk}{2!|\bs D|^3} 
\bigl[
-\gd_{nk}\gd_{ij} + \gd_{nj}\gd_{ik} + \gd_{ni}\gd_{jk}
\\&&\notag
+3\times\bigl(
\\&&\notag
\hat{D}_n\hat{D}_k\gd_{ij} -\hat{D}_n\hat D_j\gd_{ik}-\hat{D}_n\hat D_i\gd_{jk}-
\\ &&\notag
\hat{D}_i \hat{D}_j\gd_{nk}   -\hat{D}_k \hat{D}_j\gd_{ni}  -\hat{D}_i \hat{D}_k\gd_{nj}  \bigr) 
\\&&
+3\times 5\times\hat{D}_n\hat{D}_k\hat{D}_i \hat{D}_j
\bigr].
\ee

\paragraph*{Third order derivatives.--}
Similarly we find for the third order derivatives 
\be
H_{ij,knl}
&:=&\notag
\f{1}{3!}\p_l\p_n\p_kH_{ij}
\ee
the explicit representation
\be
H_{ij,knl}
&=&\notag
\f{\gk}{3!|\bs D|^4}\biggl\{
\\&&\notag
3\times \bigl[
\hat{D}_l  (\gd_{nk}\gd_{ij} -\gd_{nj}\gd_{ik} - \gd_{ni}\gd_{jk})+
\\&&\notag
\hat{D}_n (\gd_{lk}\gd_{ij}-\gd_{lj}\gd_{ik}-\gd_{li}\gd_{jk}) +
\\&&\notag
\hat{D}_k(\gd_{ln}\gd_{ij} -\gd_{lj} \gd_{ni}  -\gd_{li} \gd_{nj} )-
\\&&\notag
\hat D_i(\gd_{ln}\gd_{jk} + \gd_{lj}\gd_{nk}   +\gd_{lk} \gd_{nj})   -
\\&&\notag
\hat{D}_j(\gd_{lk}\gd_{ni} + \gd_{ln}\gd_{ik} + \gd_{li}\gd_{nk} ) \bigr]
\\&&\notag
+3\times 5\times\bigl(
\\ &&\notag
-\hat{D}_l\hat{D}_n\hat{D}_k\gd_{ij} +\hat{D}_l\hat{D}_n\hat D_j\gd_{ik}+
\\ &&\notag
\hat{D}_l\hat{D}_n\hat D_i\gd_{jk} + 
\hat{D}_l\hat{D}_i \hat{D}_j\gd_{nk}   +
\\&&\notag
\hat{D}_l\hat{D}_k \hat{D}_j\gd_{ni}  +\hat{D}_l\hat{D}_i \hat{D}_k\gd_{nj}   +
\\&&\notag
\hat{D}_k\hat{D}_i \hat{D}_j\gd_{ln}+
\hat{D}_n\hat{D}_i \hat{D}_j\gd_{lk}+
\\&&\notag
\hat{D}_n\hat{D}_k \hat{D}_j \gd_{li}+
\hat{D}_n\hat{D}_k \hat{D}_i\gd_{lj} \bigr) 
\\&&
-3\times 5\times 7\hat{D}_l\hat{D}_n\hat{D}_k\hat{D}_i \hat{D}_j
\biggr\}.
\ee
\paragraph*{Fourth order partial derivatives.--}
Finally, the fourth order derivatives, defined by
\be
H_{ij,knlm}
&:=&\notag 
\f{1}{4!}\p_m\p_l\p_n\p_kH_{ij}
\ee
are obtained as
\be
H_{ij,knlm}
&=&\notag
\f{\gk}{4!|\bs D|^5}
\biggl\{3\times \bigl[
\\&&\notag
\gd_{ml}(\gd_{nk}\gd_{ij} -\gd_{nj}\gd_{ik} - \gd_{ni}\gd_{jk})
\\&&\notag
+\gd_{mn}(\gd_{lk}\gd_{ij}-\gd_{lj}\gd_{ik}-\gd_{li}\gd_{jk})
\\&&\notag
+\gd_{mk}(\gd_{ln}\gd_{ij} -\gd_{lj} \gd_{ni}  -\gd_{li} \gd_{nj} )
\\&&\notag
-\gd_{mi}(\gd_{ln}\gd_{jk} + \gd_{lj}\gd_{nk}   +\gd_{lk} \gd_{nj})  
\\&&\notag
-\gd_{mj}(
\gd_{lk}\gd_{ni} + \gd_{ln}\gd_{ik} + \gd_{li}\gd_{nk} )\bigr]
\\&&\notag
+
\,3\times 5 \times\bigl[
\\&&\notag
\hat{D}_m\hat D_i( \gd_{ln}\gd_{jk} + \gd_{lj}\gd_{nk}   +\gd_{lk} \gd_{nj})   
\\&&\notag
+ \hat{D}_m\hat{D}_j(\gd_{lk}\gd_{ni} + \gd_{ln}\gd_{ik} + \gd_{li}\gd_{nk} ) 
\\&&\notag
+\hat{D}_l\hat D_i( \gd_{mn}\gd_{jk} + \gd_{mj}\gd_{nk}   +\gd_{mk} \gd_{nj})   
\\&&\notag
+\hat{D}_l\hat D_j( \gd_{mn}\gd_{ik} + \gd_{mi}\gd_{nk}   +\gd_{mk} \gd_{ni})   
\\&&\notag
+\hat{D}_n\hat D_i( \gd_{ml}\gd_{jk} + \gd_{mj}\gd_{lk}   +\gd_{mk} \gd_{lj})   
\\&&\notag
+\hat{D}_n\hat D_j( \gd_{ml}\gd_{ik} + \gd_{mi}\gd_{lk}   +\gd_{mk} \gd_{li})   
\\&&\notag
+\hat{D}_k\hat D_i( \gd_{ml}\gd_{jn} + \gd_{mj}\gd_{ln}   +\gd_{mn} \gd_{lj})   
\\&&\notag
+\hat{D}_k\hat D_j( \gd_{ml}\gd_{in} + \gd_{mi}\gd_{ln}   +\gd_{mn} \gd_{li})   
\\&&\notag
+\hat{D}_i\hat D_j( \gd_{ml}\gd_{kn} + \gd_{mk}\gd_{ln}   +\gd_{mn} \gd_{lk})   
\\&&\notag
-\hat{D}_m\hat{D}_l (\gd_{nk}\gd_{ij} -\gd_{nj}\gd_{ik} - \gd_{ni}\gd_{jk})
\\&&\notag
- \hat{D}_m\hat{D}_n (\gd_{lk}\gd_{ij}-\gd_{lj}\gd_{ik}-\gd_{li}\gd_{jk})  
\\&&\notag
- \hat{D}_m\hat{D}_k(\gd_{ln}\gd_{ij} -\gd_{lj} \gd_{ni}  -\gd_{li} \gd_{nj} )
\\&&\notag
- \hat{D}_n\hat{D}_k(\gd_{ml}\gd_{ij} - \gd_{mi}\gd_{lj} - \gd_{li}\gd_{mj} ) 
\\&&\notag
- \hat{D}_l\hat{D}_k(\gd_{mn}\gd_{ij} - \gd_{mi}\gd_{nj} - \gd_{ni}\gd_{mj} ) 
\\&&\notag
- \hat{D}_l\hat{D}_n(\gd_{mk}\gd_{ij} - \gd_{mi}\gd_{jk} - \gd_{ik}\gd_{mj} ) 
\bigr]
\\&&\notag
+\;3\times5\times7\times \bigl[
\\\notag
&&
\hat{D}_m\hat{D}_l\hat{D}_n\hat{D}_k\gd_{ij} 
-\hat{D}_m\hat{D}_l\hat{D}_n\hat D_j\gd_{ik}
\\&&\notag
-\hat{D}_m\hat{D}_l\hat{D}_n\hat D_i\gd_{jk} 
-\hat{D}_m\hat{D}_l\hat{D}_i \hat{D}_j\gd_{nk}  
\\&&\notag
-\hat{D}_m\hat{D}_l\hat{D}_k \hat{D}_j\gd_{ni}  
-\hat{D}_m\hat{D}_l\hat{D}_i \hat{D}_k\gd_{nj}  
\\&&\notag
-\hat{D}_m\hat{D}_k\hat{D}_i \hat{D}_j\gd_{ln}
-\hat{D}_m\hat{D}_n\hat{D}_i \hat{D}_j\gd_{lk}
\\&&\notag
-\hat{D}_m\hat{D}_n\hat{D}_k \hat{D}_j \gd_{li}
-\hat{D}_m\hat{D}_n\hat{D}_k \hat{D}_i\gd_{lj} 
\\&&\notag
-\gd_{ml}\hat{D}_n\hat{D}_k\hat{D}_i \hat{D}_j
-\gd_{mn}\hat{D}_l\hat{D}_k\hat{D}_i \hat{D}_j
\\&&\notag
-\gd_{mk}\hat{D}_l\hat{D}_n\hat{D}_i \hat{D}_j
-\gd_{mi}\hat{D}_l\hat{D}_n\hat{D}_k \hat{D}_j
\\&&\notag
-\gd_{mj}\hat{D}_l\hat{D}_n\hat{D}_k\hat{D}_i 
\bigr] 
\\&&\notag
+3\times 5\times 7\times 9\times
 \hat{D}_m\hat{D}_l\hat{D}_n\hat{D}_k\hat{D}_i \hat{D}_j
\biggr\}
\ee

\bibliographystyle{elsarticle-num}
\bibliography{numerics,StochResonance,S_Cilia,S_Experiments,S_Hydro,S_Julia,S_Particle,S_Reviews,S_BrownianSwimmer,S_Polymer_Hydro,Journals,Hanggi,BrownianMotors,GPU_examples,S_Dumbbell,S_Field}

\end{document}